\documentclass[10pt, conference, letterpaper]{IEEEtran} 
\usepackage{amsfonts}
\usepackage{mathrsfs}
\usepackage{amssymb}
\usepackage{wrapfig}
\usepackage{amsmath}
\usepackage{accents}
\usepackage{dsfont}
\usepackage{amsbsy}
\usepackage{setspace}
\usepackage{graphicx}
\usepackage{subfig}
\usepackage{url}
\usepackage{color}
\usepackage{epstopdf}
\usepackage{amssymb}
\usepackage{framed}
\usepackage{bbm}
\PassOptionsToPackage{bookmarks={false}}{hyperref}

\def\argmax{\operatornamewithlimits{arg\,max}}
\def\argmin{\operatornamewithlimits{arg\,min}}

\begin{document}
\title{Optimal Load-Balancing for High-Density Wireless Networks with Flow-Level Dynamics
\author{\IEEEauthorblockN{Bin Li$^\ast$, Xiangqi Kong$^\ast$ and Lei Wang$^\dag$}
\IEEEauthorblockA{$^\ast$Department of Electrical, Computer and Biomedical Engineering, University of Rhode Island, USA}
\IEEEauthorblockA{$^\dag$School of Software, Dalian University of Technology, China}
\IEEEauthorblockA{Email: binli@uri.edu, xqkong@my.uri.edu, lei.wang@dlut.edu.cn}
}}



\date{}

\maketitle
\newtheorem{theorem}{Theorem}
\newtheorem{lemma}{Lemma}
\newtheorem{claim}{Claim}
\newtheorem{proposition}{Proposition}
\newtheorem{corollary}{Corollary}
\newtheorem{definition}{Definition}
\newtheorem{assumption}{Assumption}
\newtheorem{remarks}{Remarks}
\newtheorem{algorithm}{Algorithm}

\newcommand{\cS}{\mathcal{S}}
\newcommand{\vS}{\mathbf{S}}
\newcommand{\vQ}{\mathbf{Q}}
\newcommand{\bE}{\mathds{E}}
\newcommand{\mc}{\mathcal}
\newcommand{\mb}{\mathbf}
\newcommand{\bs}{\boldsymbol}
\newcommand{\ol}{\overline}
\newcommand{\wt}{\widetilde}

\makeatletter
\newcommand{\rmnum}[1]{\romannumeral #1}
\newcommand{\Rmnum}[1]{\expandafter\@slowromancap\romannumeral #1@}
\makeatother
\def\maximize{\operatornamewithlimits{Maximize}}
\def\limitsup{\operatornamewithlimits{limsup}}
\def\limitinf{\operatornamewithlimits{liminf}}

\begin{abstract}
We consider the load-balancing design for forwarding incoming flows to access points (APs) in high-density wireless networks with both channel fading and flow-level dynamics, where each incoming flow has a certain amount of service demand and leaves the system once its service request is complete (referred as \emph{flow-level dynamic model}). The efficient load-balancing design is strongly needed for supporting high-quality wireless connections in high-density areas. Despite the presence of a variety of earlier works on the design and analysis of the load-balancing schemes in wireless networks, there does not exist a work on the load-balancing design in the realistic flow-level dynamic model. 

In this work, we propose a Join-the-Least-Workload (JLW) Algorithm that always forwards the incoming flows to the AP with the smallest workload in the presence of flow-level dynamics. However, our considered flow-level dynamic model differs from traditional queueing model for wireless networks in the following two aspects: (1) the dynamics of the flows is \emph{short-term} and flows will leave the network once they received the desired amount of service; (2) each individual flow faces an independent channel fading. These differences pose significant challenges on the system performance analysis. To tackle these challenges, we perform Lyapunov-drift-based analysis of the stochastic network taking into account sharp flow-level dynamics. Our analysis reveals that our proposed JLW Algorithm not only achieves maximum system throughput, but also minimizes the total system workload in heavy-traffic regimes. Moreover, we observe from both our theoretical and simulation results that the mean total workload performance under the proposed JLW Algorithm does not degrade as the number of APs increases, which is strongly desirable in high-density wireless networks. 
\end{abstract}

\section{Introduction}

With the rapid growth of smart phones, there is a strong need for high-quality wireless local access network (WLAN) connections in high-density areas, such as convention centers, auditoriums, hotel meeting rooms, lecture halls, sports stadiums, and concert halls. These high-speed wireless connections are not only for business and entertainment purposes, but more importantly provide emergency response communications in crowded places in response to unexpected events such as fire, shooting, and terrorist attack. To support such wireless connections in high-density WLANs, multiple access points (APs) are necessary to be deployed for providing satisfactory services for wireless users. 
However, in conventional WLANs, each user is automatically associated with the AP that has the best channel quality, which causes significant load imbalance among APs and results in poor network performance (e.g. \cite{kotz2005analysis}). This raises a natural question in how to develop an efficient joint load-balancing and scheduling algorithm that first determines which AP an incoming user should associate with, then each individual AP needs to decide which users it serves. The goal of such an algorithm is to maximize system throughput (or equivalently support network users as many as possible) and to minimize average user's delay.

While load-balancing for multiple APs with various fairness criteria (e.g., \cite{bejerano2004fairness,gong2008dynamic,li2014ap,
athanasiou2015optimizing,dwijaksara2016joint,sun2017novel}) have been studied extensively, relatively limited work on the realistic model exists where a mobile user transmits data from a file (or a flow), and either departs or becomes silent for a while, which was observed in prior work (e.g., \cite{balachandran2002characterizing,kotz2005analysis}). In such practical wireless networks, even the design of scheduling algorithms in a single AP case is quite non-trivial, let alone load-balancing design. Indeed, most existing scheduling designs including the well-known MaxWeight-type algorithms (e.g., \cite{taseph92,taseph93}) implicitly assume that 
the system consists of a fixed number of persistent users that continuously inject packets into the network and will never leave the network, and thus perform poorly in the presence of dynamic flows (e.g., \cite{borst09, venbsying13}). The main reason is that the queue-length-based MaxWeight algorithm myopically selects a feasible schedule with the maximal residual size of dynamic flows and hence the flows with small backlogs may stay in the network forever. Subsequent works (e.g. \cite{yingsrikant10,yingsrikantsig10,sadiq09}) have developed throughput-optimal scheduling algorithms that do not require any prior knowledge of channels and user demands. Despite these advances in efficient scheduling design for wireless networks with flow-level dynamics, the load-balancing design among multiple APs is far less explored. 

On the contrary, the load-balancing schemes have been explored extensively in data centers that distribute arriving jobs across servers with the goal of minimizing queueing delays. The celebrated Join-the-Shortest-Queue (JSQ) policy (e.g., \cite{whi86}, \cite{foschini1978basic}), where all arrivals are forwarded to the shortest queue, has been shown to not only achieve maximum system throughput but also minimize mean delay in the symmetric case \cite{taseph93}, or under the heavy-traffic regime (e.g., \cite{foschini1978basic}, \cite{erysri12}). There are many variants of JSQ policy, such as Join-the-Least-Loaded-Queue (JLQ) policy (e.g., \cite{gupta2007analysis}) instead forwarding incoming jobs to the queue with the smallest amount of remaining work (or workload), and low-communication overhead load-balancing schemes (e.g., Power-of-Two-Choices \cite{vvedenskaya1996queueing,mitzenmacher2001power}, Batch Sampling \cite{ousterhout2013sparrow,ying2015power}). However, all these load-balancing schemes presume that queueing disciplines are either First-Come-First-Serve (FCFS) (e.g., \cite{foschini1978basic,whi86,vvedenskaya1996queueing,
mitzenmacher2001power,ousterhout2013sparrow,ying2015power}) or Processor-Sharing (PS) (e.g., \cite{bonomi1990job,gupta2007analysis}), and their performance is unclear in wireless networks with both channel fading and flow-level dynamics, where each individual flow (or job) faces an independent wireless fading channel. This motivates us to investigate efficient load-balancing design in the presence of flow-level dynamics with wireless fading. The following list highlights our contributions as well as the outline of the remainder of the paper:

$\bullet$ In Section \ref{sec:model}, we formulate the problem of load-balancing design among multiple APs in high-density wireless networks in the presence of flow-level dynamics. 

$\bullet$ In Section \ref{sec:existing}, we present two existing policies and show their performance deficiencies.

$\bullet$ In Section \ref{sec:JLW}, we propose an efficient load-balancing scheme for wireless networks with flow-level dynamics, and show that it not only achieves maximum system throughput but also minimizes the mean system workload in heavily loaded conditions.

$\bullet$ We support our analytical results with extensive simulations in Section \ref{sec:simulation}, which not only confirm our theoretical findings but also exhibit the excellent performance of our proposed algorithm in general cases.

A note on Notation: We use bold and script font of a variable to denote
a vector and a set, respectively.
We use $\langle\mb{x},\mb{y}\rangle$ to denote the inner product
of two vectors $\mb{x}$ and $\mb{y}$. Let $\|\mb{x}\|_{1}$ and $\|\mb{x}\|$ denote the $l_1$ and $l_2$ norm of the vector $\mb{x}$, respectively. We also use $\mb{x}\succeq\mb{y}$ to denote that each component of vector $\mb{x}$ is not less than that of vector $\mb{y}$.

\section{System Model}
\label{sec:model}
We consider a wireless network with $M$ access points (APs). We assume that the system operates in a \emph{slotted time} manner. Here, we assume that these $M$ APs operate in orthogonal channels and can serve users (referred as flows in the rest of the paper) at the same time. However, within each AP, due to the wireless interference, at most one flow can be served in each time slot. 

Let $A_{\Sigma}[t]$ denote the number of flows arriving at the system in time slot $t$ that is independently and identically distributed (i.i.d.) over time with mean $\lambda_{\Sigma}>0$, and $A_{\Sigma}[t]\leq A_{\max}$ for some positive number $A_{\max}$, $\forall t\geq0$. We use $F_{j}[t]$ to denote the number of packets of newly arriving flow $j$ that follows any probability distribution with mean $\eta>0$, and $F_{j}[t]\leq F_{\max}$ for some $0<F_{\max}<\infty$, $\forall t\geq0$. We use $N_{m}[t]$ to denote the number of flows in AP $m$ in time slot $t$. We also use $\mc{A}_{\Sigma}[t]$ and $\mc{N}_{m}[t]$ to denote the set of newly arriving flows at the system and the set of existing flows in AP $m$ in time slot $t$, respectively. Let $R_{m,j}[t]$ be the number of residual packets of flow $j$ in AP $m$ in time slot $t$. 

We assume that each AP has $K+1$ possible channel rates $c_0, c_1,c_2,\ldots,c_K$ with $0=c_0<c_1<c_2<\ldots<c_K=c_{\max}$, where $c_k$ is a positive integer number denoting that at most $c_k$ packets can be delivered in one time slot, $\forall k=1,2,\ldots,K$. We use $C_{m,j}[t]$ to capture wireless channel fading of each flow $j$ in the $m^{\text{th}}$ AP, which measures the maximum number of packets that can be transmitted in time slot $t$ if flow $j$ is scheduled in time slot $t$. We assume that $(C_{m,j}[t])_{j\in\mc{N}_{m}[t]}$ are independently distributed across APs and
i.i.d. over both time and flows within each AP with probability distribution
\begin{align}
\Pr\left\{C_{m,j}[t]=c_k\right\}=p_{m,k}, \forall k=0,1,2,\ldots,K.
\end{align}
Here, we reasonably assume that both probability that the channel for each flow is unavailable and achieves the maximum channel rate are strictly positive, i.e., $p_{m,0}>0$ and $p_{m,K}>0$, $\forall m=1,2,\ldots,M$.

In order to characterize the underlying dynamics of flows, we introduce following notations. Let $W_m[t]\triangleq\sum_{j\in\mc{N}_{m}[t]}\left\lceil R_j[t]/c_{\max}\right\rceil$ be the total workload in AP $m$ in time slot $t$ that measures the minimum number of slots required for completing all existing service requests in AP $m$. We use $\nu_{\Sigma}[t]\triangleq\sum_{j\in\mc{A}_{\Sigma}[t]}\lceil F_j[t]/c_{\max}\rceil$ and $\nu_{m}[t]\triangleq\sum_{j\in\mc{A}_{m}[t]}\left\lceil F_j[t]/c_{\max}\right\rceil$ to denote the total amount of new workload arriving at the system and the amount of new workload injected to AP $m$ under some load-balancing policy in time slot $t$, respectively, where $\mc{A}_{m}[t]$ denotes the set of arriving flows at AP $m$ in time slot $t$. We also use $A_m[t]$ to represent the number of newly arriving flows at AP $m$ in time slot $t$. Let $\rho\triangleq\bE[\nu_{\Sigma}[t]]=\lambda_{\Sigma}w$ be the traffic intensity, where $w\triangleq\bE\left[\lceil F_j[t]/c_{\max}\rceil\right]$ denotes the expected minimum number of slots required for serving a newly arriving flow.

We define $\mu_m[t]$ to be the amount of workload decreasing at AP $m$ in time slot $t$. Since the maximum channel rate is $c_{\max}$, $\mu_{m}[t]$ is equal to either $0$ or $1$. In addition, if at least one of flows in AP $m$ has the maximum channel rate $c_{\max}$ in time slot $t$, then $\mu_m[t]=1$. Therefore, $\mu_{m}[t]\geq\mathbbm{1}_{\mc{F}_m}$, where $\mc{F}_m$ denotes the event that at least one of flows in AP $m$ has the maximum channel rate $c_{\max}$. Based on the above setup, the evolution of the workload $W_m[t]$ at each AP $m$ can be described as follows:
\begin{align}
\label{eqn:dynamic:workload}
W_m[t+1] = W_m[t] + \nu_m[t] - \mu_m[t], \forall m=1,\ldots,M.
\end{align}

We call AP $m$ \emph{stable} if its average workload is finite, i.e., 
\begin{align*}
\limsup_{T\rightarrow\infty}\frac{1}{T}\sum_{t=0}^{T-1}\bE\left[W_{m}[t]\right]<\infty.
\end{align*}
We say that the system is stable if all its APs are stable. The \emph{capacity region} $\Lambda$ is defined as a maximum set of traffic intensity $\rho$ for which the system is stable under some policy. It is shown in Appendix \ref{APP:capacity} that $\Lambda=\{\rho: \rho\leq M\}$, where we recall that $M$ is the number of APs. Note that $\rho$ denotes the average amount of incoming workload, which is the expected minimum number of slots required for serving incoming flows. On the other hand, $M$ is the maximum amount of workload that can decrease in each time slot. In order to make the system stable, $\rho$ should not be greater than $M$.

The \emph{throughput-optimal} algorithm stabilizes the system for any traffic intensity lying strictly inside the capacity region $\Lambda$. In this paper, we focus on the performance of load-balancing schemes that determine which AP should serve the newly incoming flows. We are interested in a load-balancing scheme for high-density wireless networks that not only support flows as many as possible, but also complete service requests of existing flows as fast as possible. The first goal is equivalent to maximizing system throughput, while the second goal can be achieved by minimizing the total mean system workload that measures the expected minimum number of time slots to finish all existing service requests. In the rest of the paper, we consider the following scheduling policy (see \cite{borst09}) within each AP: in each time slot, each AP $m$ always serves a flow $j^{*}_{m}$ with the maximum channel rate among all its existing flows, breaking ties uniformly at random, i.e., $j^{*}_{m}\in\argmax_{j\in\mc{N}_m[t]}C_{m,j}[t]$. As we show later, our proposed load-balancing algorithm together with this specific scheduling policy achieves both our desired goals.

Next, we discuss the performance deficiencies of existing policies that motivate us for further investigations.

\section{Performance Deficiencies of Existing Policies}
\label{sec:existing}
In this section, we present two existing policies and show their performance deficiencies. 

\subsection{Throughput Deficiency of Best-Channel-First Policy}
\label{sec:existing:BCF}
In this subsection, we consider the policy in conventional WLANs, where each incoming flow joins the AP with the best channel quality (e.g. \cite{kotz2005analysis}), which is given as follows:

\noindent\line(1,0){243}

\noindent\textbf{Best-Channel-First (BCF) Algorithm}: In each time slot $t$, forward each incoming flow to the AP with the largest channel rate, breaking ties uniformly at random.

\noindent\line(1,0){243}

Intuitively, more flows go to the AP with the better channel quality under the BCF Algorithm, and thus the AP with the better channel quality will be congested. This leads to the insufficient usage of APs with the relatively worst channel quality and results in the system throughput performance loss, let alone its mean workload performance. To see the throughput inefficiency of the BCF Algorithm, we consider the system with two APs, where flows at both APs face independent ON-OFF channel fading with different distributions. In particular, let $p_{m}\triangleq\Pr\{C_{m,j}[t]=1\}$ denote the probability that flow $j$ has an available channel at AP $m$, where $m=1,2$. Without loss of generality, we assume that $p_1>p_2$. For each incoming flow $j$, the probability that it will join the first AP under the BCF Algorithm is equal to 
\begin{align}
&\Pr\{C_{1,j}=1,C_{2,j}=0\}+\frac{1}{2}\Pr\{C_{1,j}=1,C_{2,j}=1\}\nonumber\\
&\qquad+\frac{1}{2}\Pr\{C_{1,j}=0,C_{2,j}=0\}=\frac{1}{2}(1+p_1-p_2).
\end{align}

\begin{figure}[htb!]
\vspace{-0.15in}
\begin{center}
\includegraphics[scale=0.5]{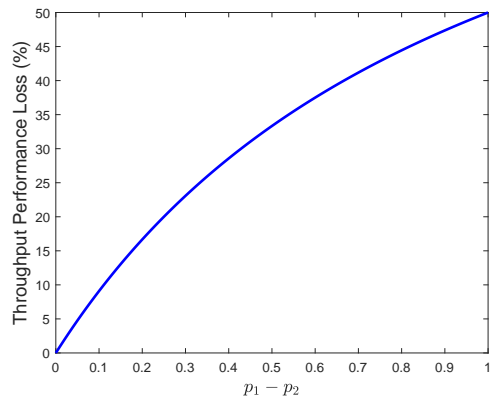}
\caption{Throughput loss under the BCF Algorithm}
\label{fig:BCF}
\end{center}
\vspace{-0.2in}
\end{figure}

Therefore, the traffic intensity to the first AP is equal to $\rho(1+p_1-p_2)/2$. In order to maintain the stability of the first AP, $\rho(1+p_1-p_2)/2\leq 1$ should be satisfied, since the workload can decrease at most by one in each AP in each time slot. Thus, the BCF Algorithm can at most support the throughput region: $\{\rho: \rho\leq 2/(1+p_1-p_2)\}$. However, the capacity region in this case is $\Lambda=\{\rho:\rho\leq 2\}$. Therefore, the BCF Algorithm suffers from throughput performance loss by $(p_1-p_2)/(1+p_1-p_2)$. For example, the throughput performance loss is $33.33\%$ when $p_1=0.9$ and $p_2=0.4$. Fig. \ref{fig:BCF} illustrates the throughput performance loss percentage with respect to the channel quality difference between two APs (i.e., $p_1-p_2$) under the BCF Algorithm. We can observe from Fig. \ref{fig:BCF} that the throughput performance loss can be as high as $50\%$ in an extreme case when the first AP always has perfect channel quality and the second AP has extremely poor channel quality, i.e., $p_1=1$ and $p_2=0$. Moreover, as the channel quality difference between two APs becomes large, the BCF Algorithm suffers from greater throughput performance loss. The reason is that if the channel quality between two APs is quite different, then the incoming flows prefer to join the AP with the better channel quality. On one hand, this results in a large number of flows accumulating at the first AP and leads to traffic congestion. On the other hand, the second AP does not have a sufficient number of flows to serve and is underutilized. In fact, a simple randomized policy that simply forwards each incoming flow to each AP uniformly at random can achieve full capacity region but it suffers from poor mean workload performance, as shown in the next subsection.

\subsection{Mean Workload Deficiency of Randomized Load-Balancing}

In this subsection, we consider both throughput and mean workload performance of a randomized load-balancing scheme, which works as follows:

\noindent\line(1,0){243}

\noindent\textbf{Randomized Load-Balancing (RLB) Algorithm}: In each time slot $t$, forward each incoming flow to each AP uniformly at random. 

\noindent\line(1,0){243}

The following proposition shows that the RLB Algorithm can indeed achieve the maximum system throughput.
\begin{proposition}
\label{prop:RLB:throughput}
The RLB Algorithm is throughput-optimal, i.e., it stabilizes the system for any traffic intensity $\rho$ lying strictly inside the capacity region $\Lambda$. Moreover, all moments of steady-state workloads are finite.
\end{proposition}
\begin{IEEEproof}
Under the RLB Algorithm, the number of incoming flows forwarded to each AP in each time slot is i.i.d. with mean $\lambda_{\Sigma}/M$ and their flow sizes are also i.i.d. with the same probability distribution as before. The proof is a special case of that in Proposition \ref{prop:JLW:throughput} in the case with a single AP and the mean flow arrival rate of $\lambda_{\Sigma}/M$, and thus is omitted for simplicity.
\end{IEEEproof}

Note that the RLB Algorithm randomly forwards the incoming flows to each AP and may cause significant workload imbalance across APs especially when the number of APs is relatively large, which is the case in high-density wireless networks. Even though this does not hurt the throughput performance, it results in the large mean workload, which implicitly degrades the mean delay performance of each flow. In order to characterize the mean workload performance, we build on the recently developed approach of using Lyapunov drifts for the steady-state analysis of queueing networks \cite{erysri12}. However, in our considered flow-level dynamic model, flows dynamically arrive at the system and will leave once they receive the desired amount of service, and existing flows suffer from independent channel fading. These two characteristics make our model quite different from the traditional FCFS queueing model that is considered in \cite{erysri12}. Thus, novel techniques are required to analyze the heavy-traffic performance of the RLB Algorithm.

To that end, we consider the workload arrival process $\{\nu_{\Sigma}^{(\epsilon)}[t]\}_{t\geq0}$, parameterized by $\epsilon>0$, with traffic intensity $\rho^{(\epsilon)}$ satisfying $\epsilon=M-\rho^{(\epsilon)}>0$ and $\text{Var}(\nu_{\Sigma}^{(\epsilon)})$. Here, $\epsilon$ characterizes the closeness of the traffic intensity to the boundary of the capacity region, and is usually referred as \emph{heavy-traffic parameter}. We are interested in understanding the steady-state workload with vanishing $\epsilon$. The next proposition shows that the RLB Algorithm results in the large mean workload even under the Bernoulli flow arrival.
\begin{proposition}
\label{prop:delay:RLB} Assume that the number of arriving flows $A_{\Sigma}[t]$ follows Bernoulli distribution. Let $\wt{\mb{W}}^{(\epsilon)}=(\wt{W}_m^{(\epsilon)})_{m=1}^{M}$ be a random vector with the same distribution as the steady-state distribution of the workload processes under the RLB Algorithm. Consider the heavy-traffic limit $\epsilon\downarrow0$, suppose that the variance $\text{Var}(\nu_{\Sigma}^{(\epsilon)})$ of the arrival process $\{\nu_{\Sigma}^{(\epsilon)}\}_{t\geq0}$ converges to a constant $\sigma^2$. Then, we have 
\begin{align}
\lim_{\epsilon\downarrow0}\epsilon\bE\left[\sum_{m=1}^{M}\wt{W}_m^{(\epsilon)}\right]=\frac{1}{2}\left(\sigma^2+M(M-1)\right).
\end{align}
\end{proposition}
\begin{IEEEproof}
Under the RLB Algorithm, the number of incoming flows joining to the $m^{\text{th}}$ AP in time slot $t$ can be represented as follows:
\begin{align}
A_{m}[t]=A_{\Sigma}[t]\mathbbm{1}_{\mc{H}[t]},
\end{align}
where $\mc{H}[t]$ denotes the event that the incoming flow joins the $m^{\text{th}}$ AP in time slot $t$,  which is independent from workload $W_m[t]$. In addition, under the RLB Algorithm, the event $\mc{H}[t]$ is i.i.d. with Bernoulli distribution with mean $1/M$. Therefore, the workload arriving at AP $m$ in time slot $t$ is 
\begin{align}
\nu_{m}[t]=\nu_{\Sigma}[t]\mathbbm{1}_{\mc{H}[t]},
\end{align}
where $\nu_{\Sigma}=A_{\Sigma}[t]\left\lceil F_j[t]/c_{\max}\right\rceil$ since $A_{\Sigma}[t]$ follows Bernoulli distribution. Thus, we have 
\begin{align}
&\bE\left[\nu_m[t]\right]=\frac{1}{M}\rho^{(\epsilon)}, \nonumber\\
\text{ and }& \text{Var}(\nu_m[t])=\frac{1}{M}\text{Var}(\nu_{\Sigma}^{(\epsilon)})+\frac{M-1}{M^2}(\rho^{(\epsilon)})^2.
\end{align}
Hence, we have 
\begin{align}
\label{eqn:prop:delay:RLB}
\lim_{\epsilon\downarrow0}\epsilon\bE\left[\wt{W}_m^{(\epsilon)}\right]=\frac{1}{2}\left(\frac{1}{M}\sigma^2+M-1\right),
\end{align}
which implies the desired result by summing over $m=1,2,\ldots,M$. The proof of \eqref{eqn:prop:delay:RLB} is a special case of that in Proposition \ref{prop:JLW:upperBound} and Proposition \ref{prop:LowerBound} in the case with a single AP and the arrival workload $\nu_{m}[t]$, and thus is omitted for conciseness.
\end{IEEEproof}

From Proposition \ref{prop:delay:RLB}, we can observe that the mean total workload under the RLB Algorithm increases quadratically with the number of APs, which is undesirable in high-density networks even when $M=10$. The main reason lies in that the RLB Algorithm randomly makes the load-balancing decision, and does not fully utilize the precious network resources. This motivates us to develop a throughput-optimal load-balancing scheme under which the mean total workload is minimized and does not suffer from performance loss as the number of APs scales. This property is pronounced in highly-dense wireless networks in the presence of many APs.

\section{Efficient Load-Balancing Design}
\label{sec:JLW}
In this section, we first propose a workload-aware load-balancing algorithm. Then, we show that the proposed algorithm not only achieves maximum system throughput but also minimizes the mean total workload in the heavy-traffic regime.  

As we discussed in the last section, the inefficiency of both BCF and RLB Algorithms lie in the fact that they are not aware of system workloads and thus cause significant load imbalance among multiple APs. This motivates us to develop a workload-aware load-balancing scheme that can evenly distribute incoming workloads across multiple APs. Motivated by the design of JSQ and JLQ polices in data centers, we propose the following workload-aware load-balancing algorithm that aims to balance workloads across multiple APs in the presence of dynamic flows.

\noindent\line(1,0){243}

\noindent\textbf{Join-the-Least-Workload (JLW) Algorithm}: In each time slot $t$, given the current workload $\mb{W}[t]=(W_m[t])_{m=1}^{M}$, forward all the arriving flows to the AP with the smallest workload, i.e., 
\begin{align}
\mb{A}^{*}[t]\in\argmin_{\mb{A}=(A_m)_{m=1}^{M}\succeq\mb{0}:\sum_{m=1}^{M}A_{m}=A_{\Sigma}[t]}\langle \mb{A},\mb{W}[t]\rangle,
\end{align}
breaking ties uniformly at random.

\noindent\line(1,0){243}

In the JLW Algorithm, the central controller broadcasts the ID of the AP with the least workload in each time slot, and thus all arriving flows will join that AP. This is possible in high-density wireless networks, where all APs are interconnected. The main difference between our JLW Algorithm and the JLQ Algorithm lies in that the JLQ Algorithm is mainly designed for the system with FCFS or PS queueing discipline, while each flow in our scenario faces an independent channel fading and potentially may have different service rate. This poses significant challenges for the performance analysis of the JLW Algorithm. Nevertheless, we can still show that the JLW Algorithm achieves maximum system throughput. Moreover, we can show that all moments of steady-state workload are bounded under the JLW Algorithm, which enables us to analyze the mean workload performance by using the Lyapunov-type approach developed in \cite{erysri12}.

\begin{proposition}
\label{prop:JLW:throughput}
The JLW Algorithm is throughput-optimal, i.e., it stabilizes the system for any traffic intensity lying strictly inside the capacity region $\Lambda$. Moreover, all moments of steady-state workloads are bounded.
\end{proposition}
\begin{IEEEproof}
The proof is available in Section \ref{sec:throughput:optimality}.
\end{IEEEproof}

Having established the throughput optimality and the moment existence of the steady-state workload of the JLW Algorithm, we are ready to analyze the mean workload performance in the heavy-traffic regime. Similar to the heavy-traffic analysis of the RLB Algorithm, we consider the arrival process $\{\nu_{\Sigma}^{(\epsilon)}[t]\}_{t\geq0}$ with heavy-traffic parameter $\epsilon>0$ characterizing the closeness of the traffic intensity to the boundary of the capacity region, i.e., $\epsilon=M-\rho^{(\epsilon)}>0$.

\begin{proposition}
\label{prop:JLW:upperBound}
Let $\wt{\mb{W}}^{(\epsilon)}=(\wt{W}_m^{(\epsilon)})_{m=1}^{M}$ be a random vector with the same distribution as the steady-state distribution of the workload processes under the JLW Algorithm. Consider the heavy-traffic limit $\epsilon\downarrow0$, suppose that the variance $\text{Var}(\nu_{\Sigma}^{(\epsilon)})$ of the arrival process $\{\nu_{\Sigma}^{(\epsilon)}\}_{t\geq0}$ converges to a constant $\sigma^2$. Then, we have 
\begin{align}
\label{eqn:prop:JLW:upperBound}
\lim_{\epsilon\downarrow0}\epsilon\bE\left[\sum_{m=1}^{M}\wt{W}_m^{(\epsilon)}\right]\leq\frac{\sigma^2}{2}.
\end{align}
\end{proposition}
\begin{IEEEproof}
The proof is available in Section \ref{sec:heavy-traffic}.
\end{IEEEproof}

In fact, the upper bound in \eqref{eqn:prop:JLW:upperBound} is also tight. To see this, we provide a generic lower bound for all feasible load-balancing policies by constructing a hypothetical single-server queue $\{\Phi[t]\}_{t\geq0}$ with the arrival process $\{\nu_{\Sigma}^{(\epsilon)}[t]\}_{t\geq0}$ and the constant service rate $M$. The queue-length evolution of this single-server queue can be described as follows: 
\begin{align}
\Phi[t+1]=\max\left\{\Phi[t]+\nu_{\Sigma}^{(\epsilon)}[t]-M,0\right\}.
\end{align}
It is easy to see that the constructed single-server queue length $\{\Phi[t]\}_{t\geq0}$ is stochastically smaller than the total workload process $\{\sum_{m=1}^{M}W_m[t]\}_{t\geq0}$ of the original system under any feasible policy, since the total system workload can at most decrease by $M$ in one time slot. Hence, by using \cite[Lemma 4]{erysri12} for the constructed single-server queue, we have the following lower bound on the steady-state workload under any feasible policy.

\begin{proposition}
\label{prop:LowerBound}
Let $\wt{\mb{W}}^{(\epsilon)}=(\wt{W}_m^{(\epsilon)})_{m=1}^{M}$ be a random vector with the same distribution as the steady-state distribution of the workload processes under any feasible load-balancing policy. Consider the heavy-traffic limit $\epsilon\downarrow0$, suppose that the variance $\text{Var}(\nu_{\Sigma}^{(\epsilon)})$ of the arrival process $\{\nu_{\Sigma}^{(\epsilon)}\}_{t\geq0}$ converges to a constant $\sigma^2$. Then,
\begin{align}
\lim_{\epsilon\downarrow0}\epsilon\bE\left[\sum_{m=1}^{M}\wt{W}_m^{(\epsilon)}\right]\geq\frac{\sigma^2}{2}.
\end{align}
\end{proposition}

This together with Proposition \ref{prop:JLW:upperBound} establishes the heavy-traffic optimality of our proposed JLW Algorithm. Moreover, compared with the mean workload performance under the RLB Algorithm in the heavy-traffic regime, the mean workload under our JLW Algorithm does not incur any performance loss by increasing the number of APs. This desirable property implies that the JLW Algorithm is suitable for deployment in high-density wireless networks.

\section{Simulation Results}
\label{sec:simulation}
In this section, we provide simulation results for our proposed JLW Algorithm and compare its performance to both BCF and RLB Algorithms. In the simulations, we assume that the number of flows arriving at the system in each time slot follows a Bernoulli distribution with mean $\lambda$. Each flow at each AP faces i.i.d. channel fading with rates $0, 1, 5, 10$ and corresponding probability $0.1, 0.2, 0.5, 0.2$. The flow size $F$ is equal to $10\times\beta$ with probability $(w-1)/(\beta-1)$ and $10$ otherwise, where we recall that $w=\bE\left[\lceil F/c_{\max}\rceil\right]$ is the mean newly arriving workload and $\beta\geq2$ is some parameter that measures the variance of the newly arriving workload. Indeed, the variance of the newly arriving workload in this setup is equal to $(w-1)\beta-w(w-1)$, which linearly increases with the parameter $\beta$. We let $w$ be equal to the number of APs $M$ and thus the capacity region $\Lambda$ is $\{\lambda:0<\lambda\leq1\}$. We set $M=5$ and $\beta=20$ in the simulations, unless we specifically mention them. 

\subsection{Throughput Performance}

\begin{figure}[!hbt]
\vspace{-0.2in}
\centerline{\subfloat[Throughput performance validation]{
\includegraphics[width=0.27\textwidth]{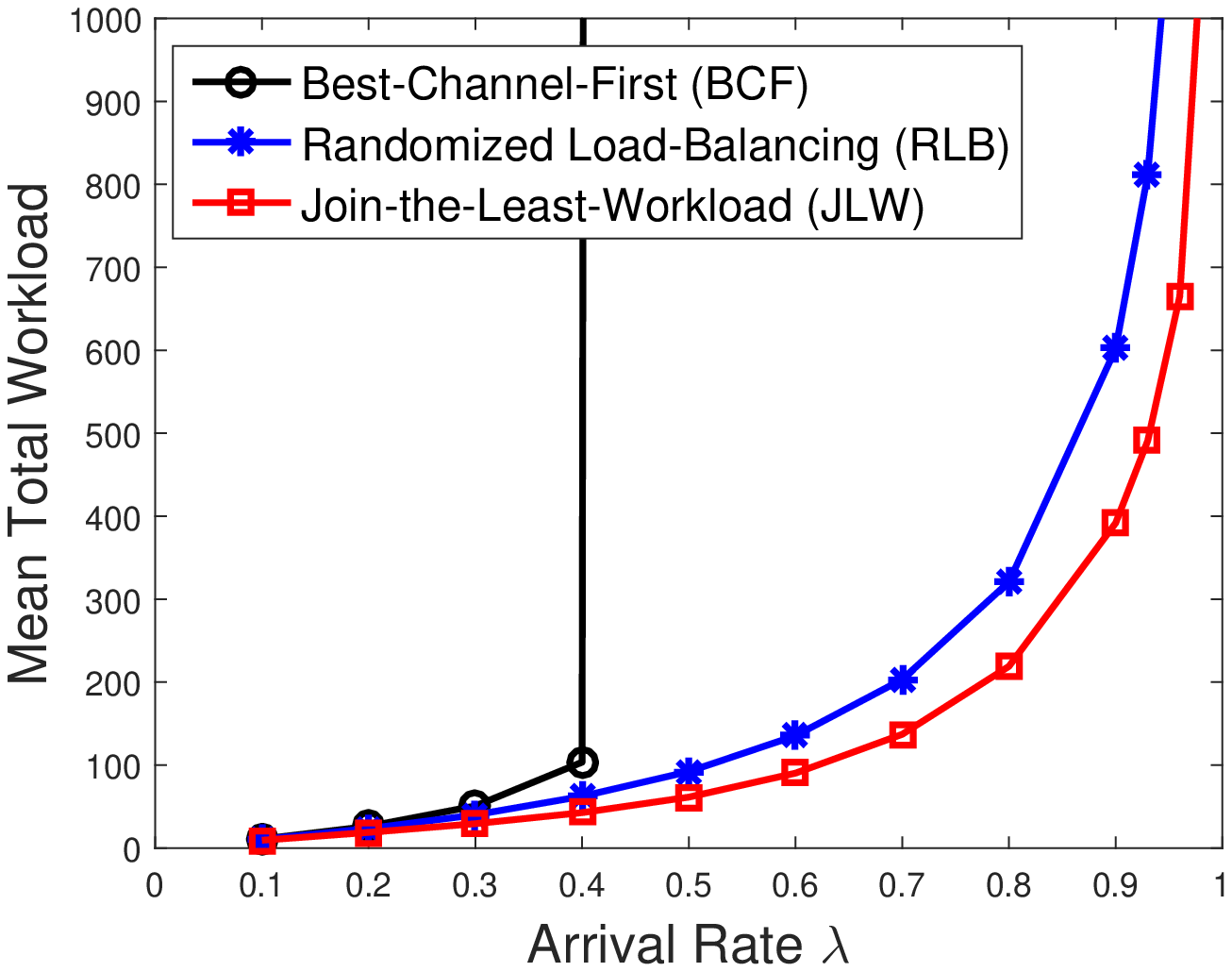}
\label{fig:throughput}} 
\hspace{-0.2in}
\subfloat[Workload reduction by JLW]{
\includegraphics[width=0.27\textwidth]{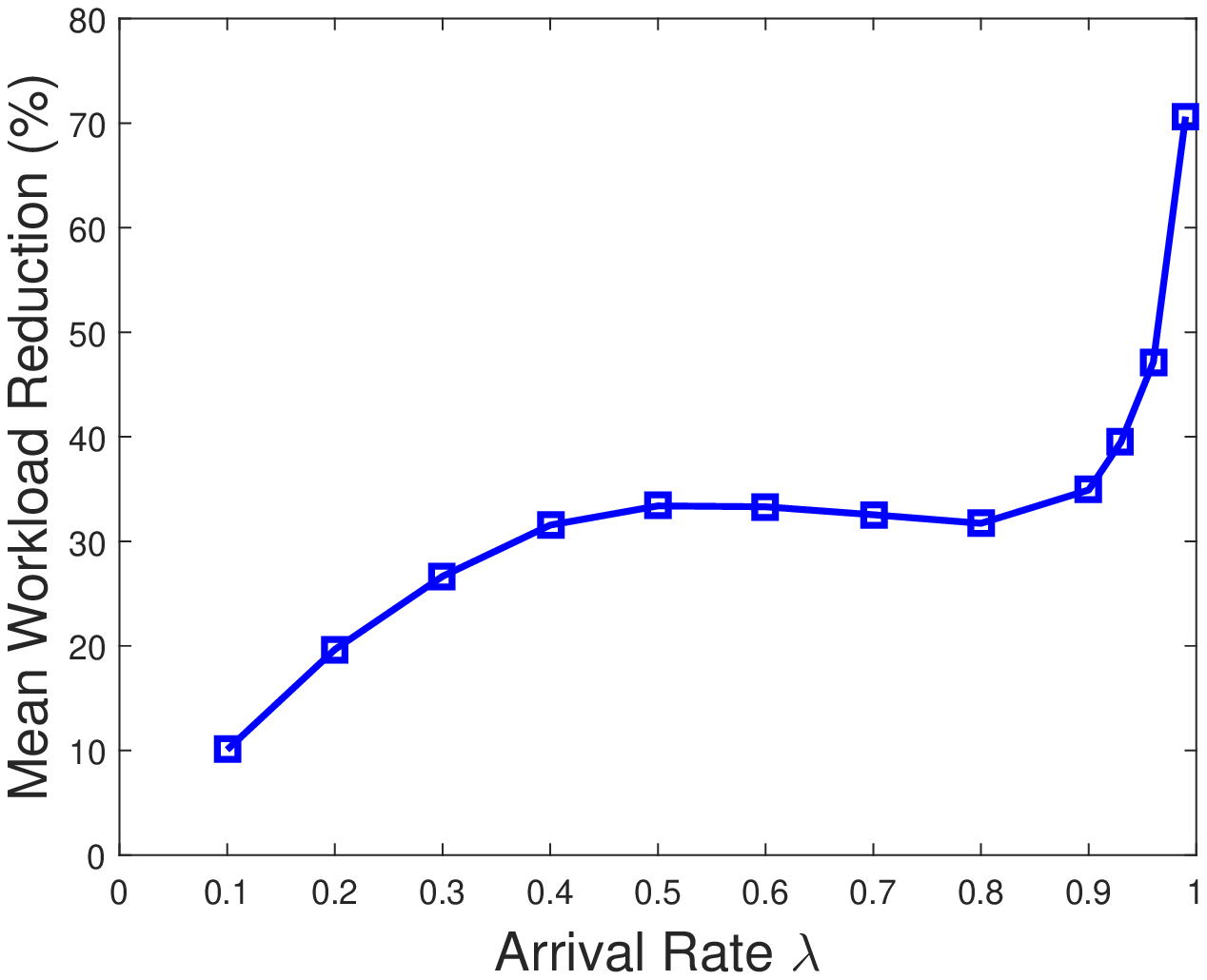}
 \label{fig:workloadReduction} } }\caption{The workload performance of the JLW Algorithm}
 \vspace{-0.2in}
\end{figure}

Fig. \ref{fig:throughput} shows the mean total workload performance versus the mean arrival rate under the BCF Algorithm, the RLB Algorithm and our proposed JLW Algorithm. We can observe from Fig. \ref{fig:throughput} that both JLW and RLB Algorithms can stabilize the system for any arrival rate $\lambda$ between $0$ and $1$, which validate their throughput optimality (cf. Proposition \ref{prop:RLB:throughput} and Proposition \ref{prop:JLW:throughput}). In contrast, the BCF Algorithm cannot support the arrival rate of $\lambda=0.45$, where the mean workload blows up. This also matches our discussions about the throughput deficiency of the BCF Algorithm in Section \ref{sec:existing:BCF}. In addition, we can see that the mean workload under the JLW Algorithm is smaller than that under the RLB Algorithm under different arrival rates. To see it more clearly, Fig. \ref{fig:workloadReduction} characterizes the mean workload reduction percentage by the JLW Algorithm compared with the RLB Algorithm. From Fig. \ref{fig:workloadReduction}, we can observe that the mean workload reduction is $10\%$ even when the arrival rate is equal to $0.1$, and can reach as high as $70\%$ when the arrival rate is $0.99$. The reason is that the RLB Algorithm randomly forwards newly arriving flows to each AP and thus may cause some APs underutilized, while the JLW Algorithm aims to balance the workloads across APs and utilizes network resources more efficiently. Thus, the JLW Algorithm shows significant performance gain over the RLB Algorithm especially when the arrival rate is high.

\subsection{Heavy-Traffic Performance}

\begin{figure}[!hbt]
\vspace{-0.2in}
\centerline{\subfloat[Heavy-traffic optimality validation]{
\includegraphics[width=0.27\textwidth]{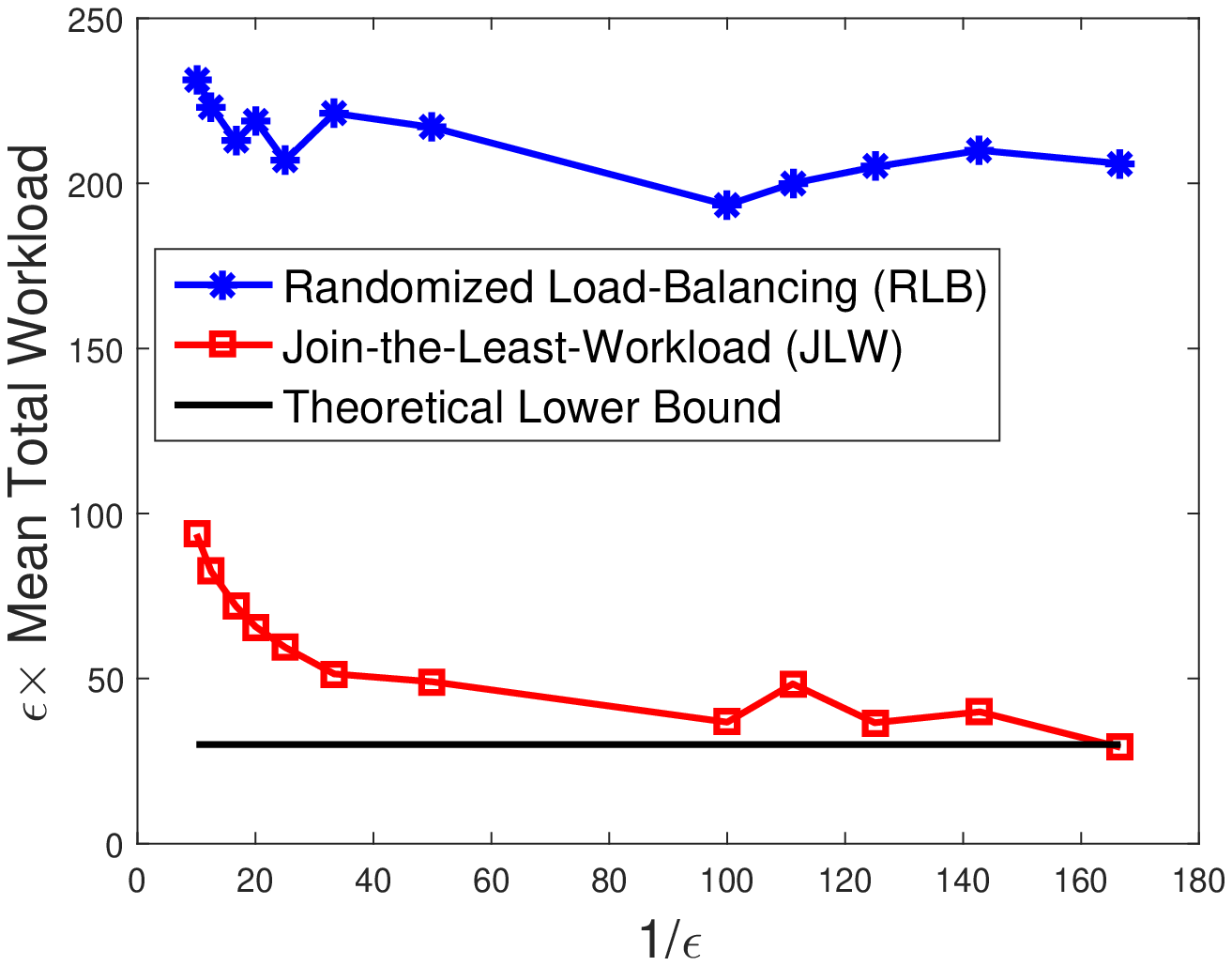}
\label{fig:heavytraffic}} 
\hspace{-0.1in}
\subfloat[Impact of number of APs]{
\includegraphics[width=0.27\textwidth]{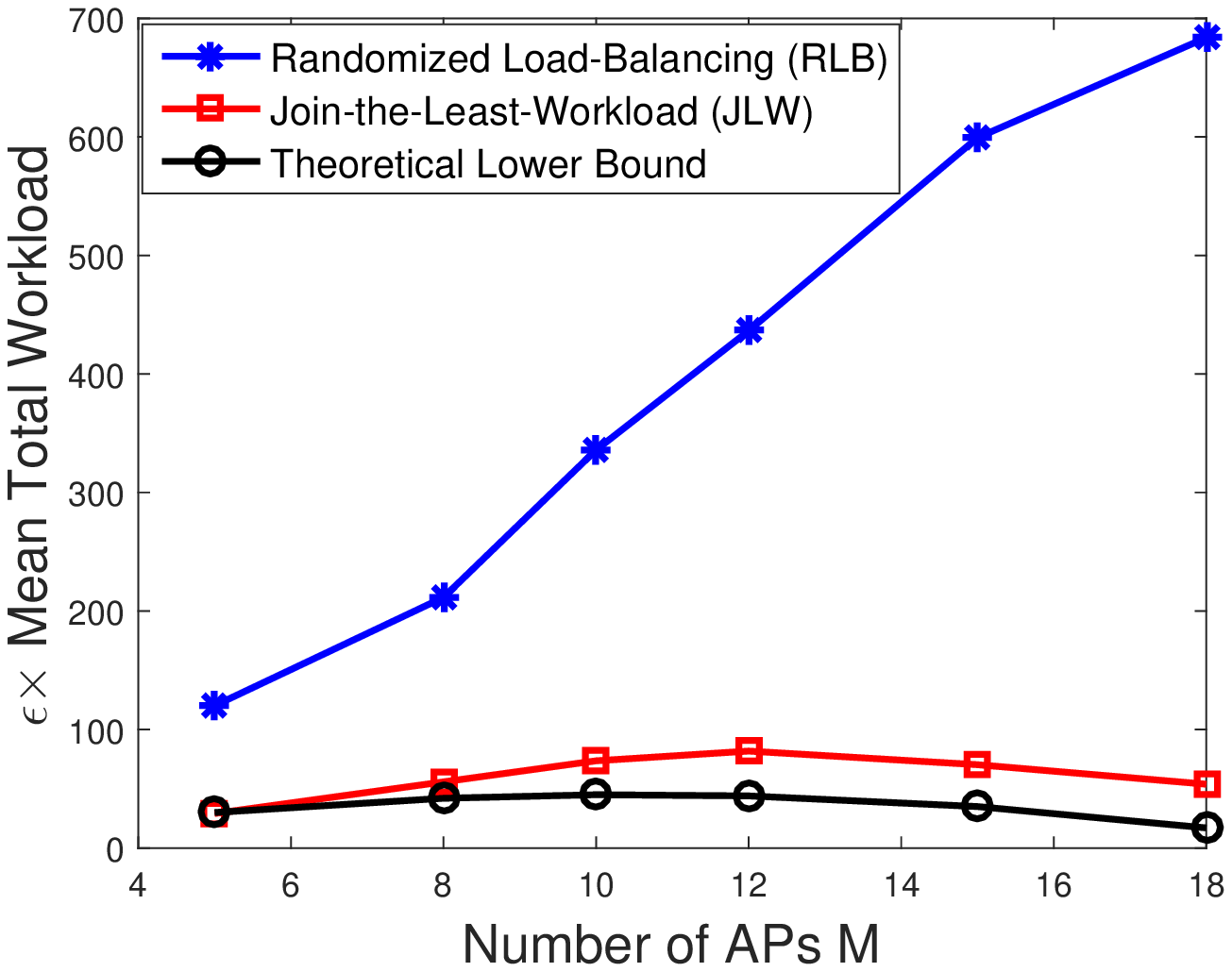}
 \label{fig:impactofNum} } }\caption{Mean workload in heavy-traffic regimes}
\vspace{-0.15in}
\end{figure}

Fig. \ref{fig:heavytraffic} shows the impact of heavy-traffic parameter $\epsilon$ on the mean total workload under both RLB and JLW Algorithms. From Fig. \ref{fig:heavytraffic}, we can observe that the mean total workload under the JLW Algorithm converges to the theoretical lower bound (equal to $30$) derived in Proposition \ref{prop:LowerBound}, while the RLB Algorithm always keeps it away from the theoretical lower bound. This confirms the heavy-traffic optimality of the JLW Algorithm, i.e., it minimizes the mean total workload as the heavy-traffic parameter $\epsilon$ diminishes.

Fig. \ref{fig:impactofNum} studies the impact of number of APs $M$ on the mean total workload under both RLB and JLW Algorithms, where we fix $\epsilon$ to $0.006$ and vary the number of APs from $5$ to $18$. From Fig. \ref{fig:impactofNum}, we can observe that the performance of the JLW Algorithm stays close to the theoretical lower bound, which is equal to $(21M-20-M^2)/2$ from Proposition \ref{prop:LowerBound} under our setting. In contrast, the product of mean total workload and heavy-traffic parameter $\epsilon$ under the RLB Algorithm is equal to $10M-10$ from Proposition \ref{prop:delay:RLB} in the heavy-traffic regime, which  linearly increases with the number of APs (also can be observed from Fig. \ref{fig:impactofNum}). This renders infeasibility of the RLB Algorithm for high-density wireless networks with many APs. 

\subsection{Mean Delay Performance}

\begin{figure}[!hbt]
\vspace{-0.2in}
\centerline{\subfloat[Comparison between RLB and JLW]{
\includegraphics[width=0.27\textwidth]{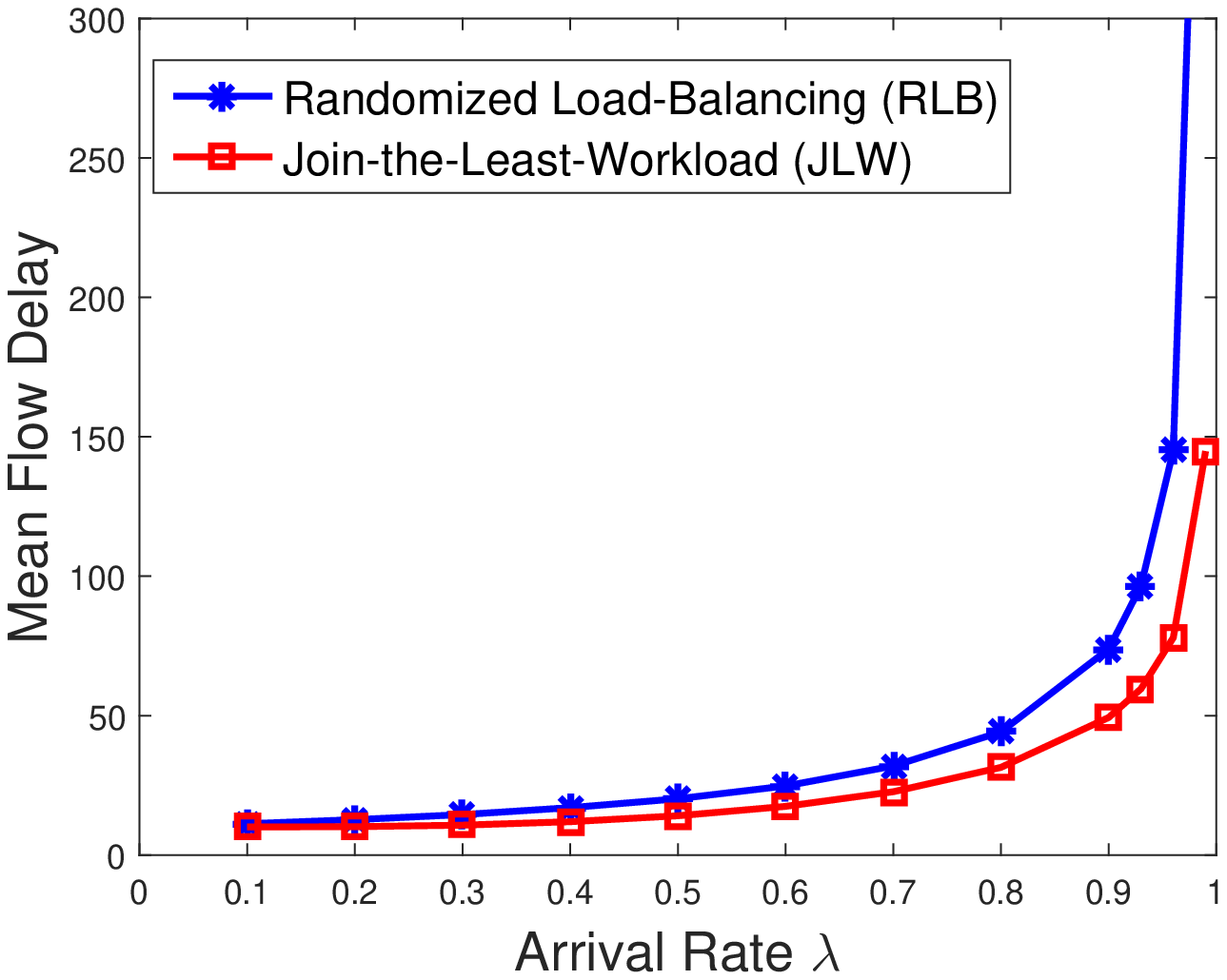}
\hspace{-0.25in}
\label{fig:delay}}\subfloat[Delay improvement by JLW]{
\includegraphics[width=0.27\textwidth]{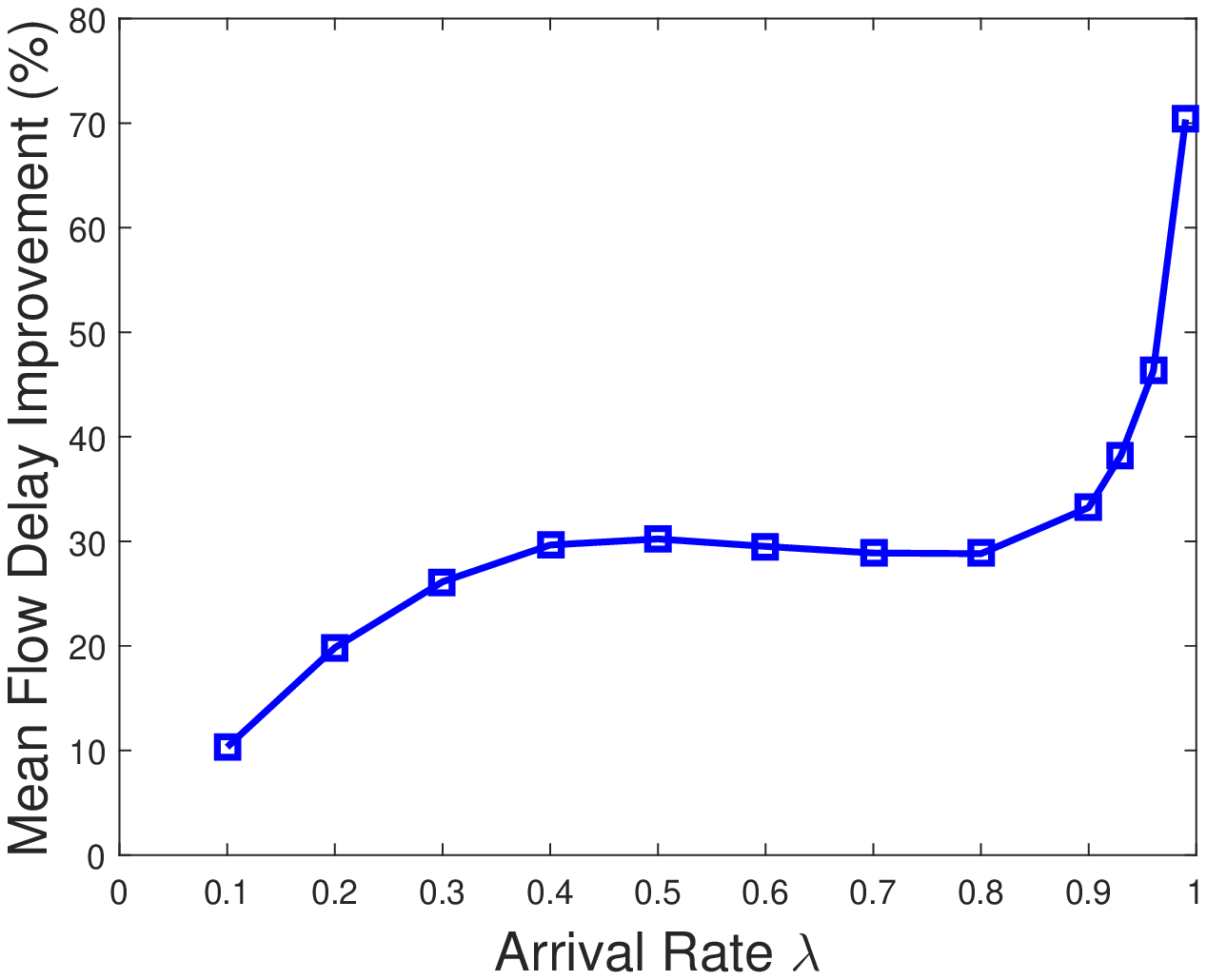}
\label{fig:delayImprovement} } }\caption{The delay performance of the JLW Algorithm}
\vspace{-0.15in}
\end{figure}

In this subsection, we study the mean delay performance of flows under both RLB and JLW Algorithms. From Fig. \ref{fig:delay}, we can observe that the JLW Algorithm outperforms the RLB Algorithm in terms of mean delay performance. Moreover, the delay improvement by the JLW Algorithm is very similar to its workload reduction compared with the RLB Algorithm. The reason lies in that the smaller workload implies that each flow spends the less waiting time in the system and thus experiences smaller delay.

\begin{figure}[!hbt]
\vspace{-0.2in}
\centerline{\subfloat[Mean delay performance]{
\includegraphics[width=0.27\textwidth]{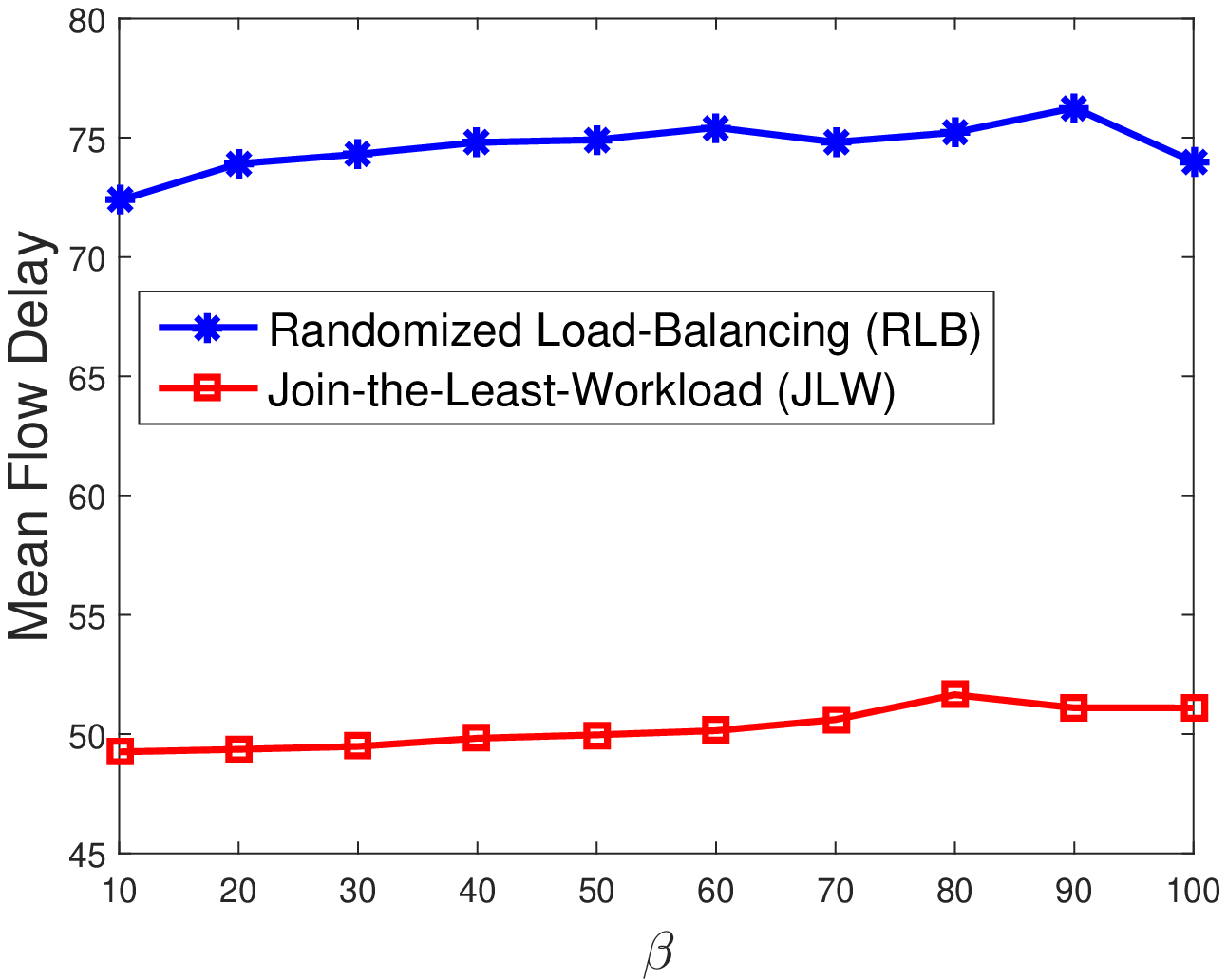}
\label{fig:delayInsensitivity}}
\hspace{-0.2in}\subfloat[Mean workload performance]{
\includegraphics[width=0.27\textwidth]{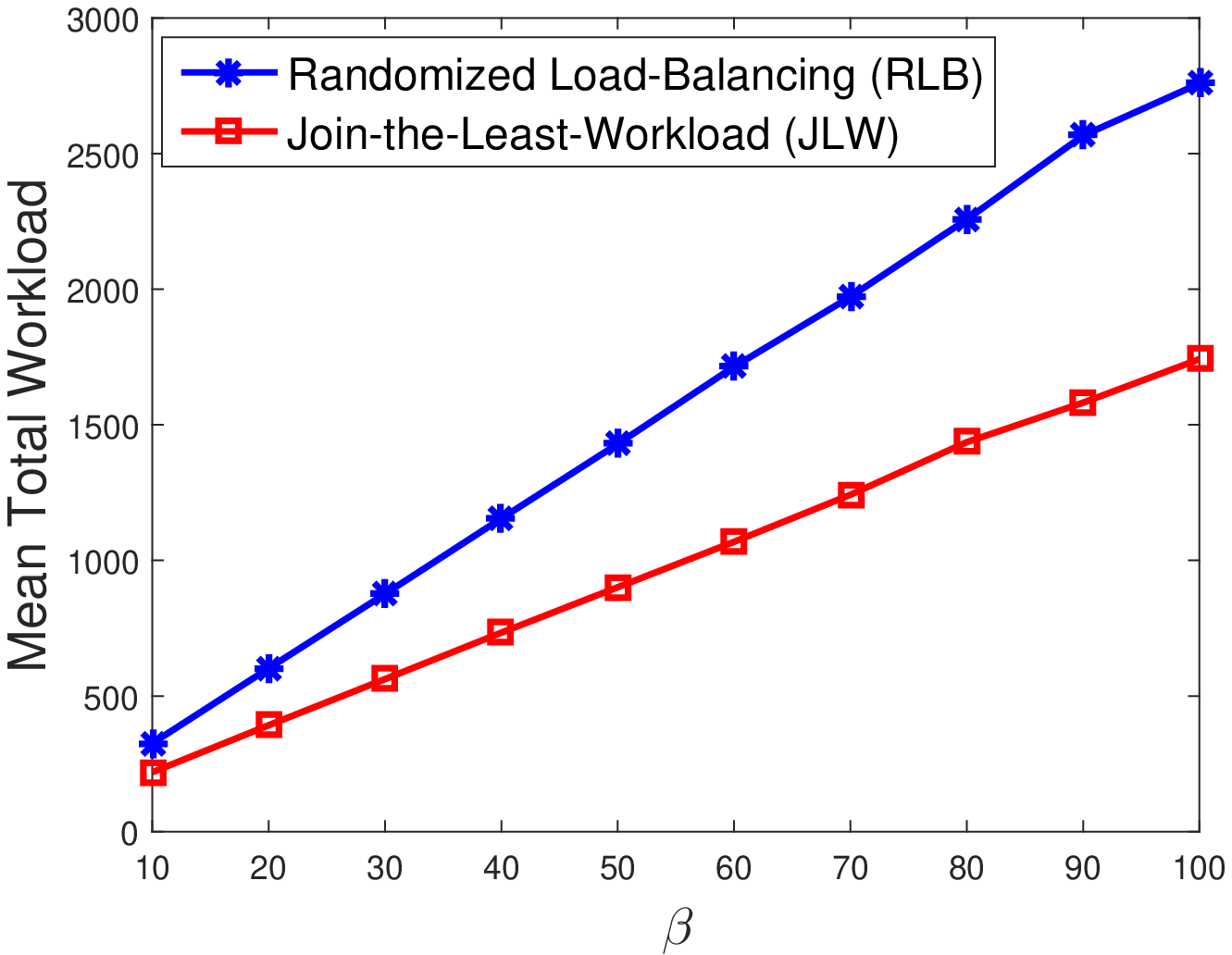}
\label{fig:workloadInsensitivity} } }\caption{The impact of the flow size distribution}
\vspace{-0.2in}
\label{fig:ImpactofDistrition}
\end{figure}

Next, we study the impact of the variance of the flow size on the system performance. Recall that the parameter $\beta$ characterizes the variance of the flow size. The larger the $\beta$, the higher variance of the flow size. Here, we fix the arrival rate $\lambda$ to $0.9$, and vary the parameter $\beta$ from $10$ to $100$. We can see from Fig. \ref{fig:delayInsensitivity} that the JLW Algorithm always performs better than the RLB Algorithm, and the parameter $\beta$ does not affect their mean delay performance. This is referred as the \emph{delay insensitivity} property in queueing literature, where the mean delay performance is insensitive to the flow-size distribution beyond its mean. This is expected, since each AP always serves the flow with the maximum channel rate, and in the extreme non-fading case, flows are served in a preemptive random order, where the mean delay performance exhibits insensitivity property to the flow size distribution (see \cite{henderson1992insensitivity}). However, different from the mean delay performance, the mean total workload under both RLB and JLW Algorithms increases linearly with the parameter $\beta$, as shown in Fig. \ref{fig:workloadInsensitivity}. The reason lies in the fact that the total system workload is lower-bounded by the queue-length of a hypothetical single-server queue $\{\Phi[t]\}_{t\geq0}$ with the new workload arrival process $\{\nu_{\Sigma}[t]\}_{t\geq0}$ and the constant service rate $M$ under any feasible load-balancing policies (also see the discussion in Section \ref{sec:JLW}), where the mean queue-length $\bE[\Phi[t]]$ is sensitive to the variance of $\nu_{\Sigma}[t]$.

\section{Throughput Optimality Analysis}

\label{sec:throughput:optimality}

In this section, we establish the throughput optimality of the JLW algorithm as well as the boundedness of all moments of steady-state workload, where the later property enables us to analyze the mean workload performance in Section \ref{sec:heavy-traffic}. We choose the Lyapunov function 
\begin{align}
V(\mb{W})\triangleq\|\mb{W}\|.
\end{align}
Then, we consider the conditional expectation of its drift $\Delta V(\mb{W})\triangleq \left(V(\mb{W}[t+1])-V(\mb{W}[t])\right)\mathbbm{1}_{\{\mb{W}[t]=\mb{W}\}}$.
\begin{align}
\label{eqn:prop:throughput:main}
&\bE\left[\Delta V(\mb{W})\middle|\mb{W}[t]=\mb{W}\right]\nonumber\\
=&\bE\left[V(\mb{W}[t+1])-V(\mb{W}[t])\middle|\mb{W}[t]=\mb{W}\right]\nonumber\\
=&\bE\left[\sqrt{\|\mb{W}[t+1]\|^2}-\sqrt{\|\mb{W}[t]\|^2}\middle|\mb{W}[t]=\mb{W}\right]\nonumber\\
\leq&\frac{1}{2\|\mb{W}\|}\bE\left[L(\mb{W}[t+1])-L(\mb{W}[t])\middle|\mb{W}[t]=\mb{W}\right],
\end{align}
where the last step is true for $L(\mb{W})\triangleq\|\mb{W}\|^2$ and follows from the fact that $f(x)=\sqrt{x}$ is concave for $x\geq0$ and thus $f(y)-f(x)\leq f'(x)(y-x)=(y-x)/(2\sqrt{x})$ with $y=\|\mb{W}[t+1]\|^2$ and $x=\|\mb{W}[t]\|^2$.

Next, we focus on the expected difference in \eqref{eqn:prop:throughput:main}, which is just the expected drift of $L(\mb{W})$. We will omit the time index $[t]$ after the first step for conciseness. 
\begin{align}
\label{eqn:prop:throughput:quadractic}
\bE\left[\Delta L(\mb{W})\middle|\mb{W}\right]=&\bE\left[\|\mb{W}[t+1]\|^2-\|\mb{W}[t]\|^2\middle|\mb{W}\right] \nonumber\\
=&\bE\left[\|\mb{W}+\bs{\nu}-\bs{\mu}\|^2-\|\mb{W}\|^2\middle|\mb{W}\right]\nonumber\\
=&\bE\left[2\langle \mb{W}, \bs{\nu}-\bs{\mu}\rangle + \|\bs{\nu}-\bs{\mu}\|^2\middle|\mb{W}\right] \nonumber\\
\leq&2\bE\left[\langle\mb{W},\bs{\nu}-\bs{\mu}\rangle\middle|\mb{W}\right] + K_1,
\end{align}
where $K_1\triangleq M(\nu_{\max}^2+1)$ is bounded and $\nu_{\max}\triangleq A_{\max}\lceil F_{\max}/c_{\max}\rceil$.

Next, we focus on $\bE\left[\langle\mb{W},\bs{\nu}-\bs{\mu}\rangle\middle|\mb{W}\right]$. Since the traffic intensity $\rho$ is strictly inside the capacity region $\Lambda$, there exists an $\epsilon$ such that $\epsilon=M-\rho>0$. We define a hypothetical arrival rate vector $\bs{\lambda}=(\lambda_m)_{m=1}^{M}$ as $\bs{\lambda}=\mb{1}/w-\epsilon\mb{1}/(wM)$, where we recall that $w=\bE\left[\lceil F_j[t]/c_{\max}\rceil\right]$. Hence, we have $\sum_{m=1}^{M}\lambda_m=M/w-\epsilon/w=\lambda_{\Sigma}$, where we use the fact that $\rho=\lambda_{\Sigma}w$. Therefore, we have 
\begin{align}
\label{eqn:prop:throughput:temp}
&\bE\left[\langle\mb{W},\bs{\nu}-\bs{\mu}\rangle\middle|\mb{W}\right]\nonumber\\
\stackrel{(a)}{=}&\langle\mb{W},w\bE\left[\mb{A}\middle|\mb{W}\right]-w\bs{\lambda}\rangle - \langle\mb{W}, \mb{1} - w\bs{\lambda}\rangle-\bE\left[\langle\mb{W}, \bs{\mu}-\mb{1}\rangle\middle|\mb{W}\right]\nonumber\\
\stackrel{(b)}{\leq}&w\langle\mb{W},\bE\left[\mb{A}\middle|\mb{W}\right]-\bs{\lambda}\rangle-\frac{\epsilon}{M}\|\mb{W}\|_{1}+\bE\left[\sum_{m=1}^{M}W_m\mathbbm{1}_{\ol{\mc{F}}_m}\middle|\mb{W}\right],
\end{align}
where step $(a)$ uses the fact that $\bE\left[\bs{\nu}\middle|\mb{W}\right]=w\bE\left[\mb{A}\middle|\mb{W}\right]$; $(b)$ uses the fact that 
$\mu_m\geq\mathbbm{1}_{\mc{F}_m}$ for all $m=1,2,\ldots,M$ and we recall that $\mc{F}_m$ denotes the event that at least one flow has the maximum channel rate $c_{\max}$ and $\ol{\mc{F}}_m$ is the complement of the event $\mc{F}_m$.

For the term $\langle\mb{W},\bE\left[\mb{A}\middle|\mb{W}\right]-\bs{\lambda}\rangle$ in \eqref{eqn:prop:throughput:temp}, we have
\begin{align}
\label{eqn:prop:throughput:temp1}
\langle\mb{W},\bE\left[\mb{A}\middle|\mb{W}\right]-\bs{\lambda}\rangle
=& W_{\min}\bE\left[A_{\Sigma}\middle|\mb{W}\right]-\langle\mb{W},\bs{\lambda}\rangle \nonumber\\
=&W_{\min}\lambda_{\Sigma}-\sum_{m=1}^{M}\lambda_mW_m\nonumber\\
=&\sum_{m=1}^{M}\lambda_m\left(W_{\min} - W_{m}\right)\nonumber\\
\leq&0,
\end{align}
where the first step follows from the definition of the JLW Algorithm.

With regard to the term $\bE\left[\sum_{m=1}^{M}W_m\mathbbm{1}_{\ol{\mc{F}}_m}\middle|\mb{W}\right]$ in \eqref{eqn:prop:throughput:temp}, we have
\begin{align}
\label{eqn:prop:throughput:temp2}
&\bE\left[\sum_{m=1}^{M}W_m\mathbbm{1}_{\ol{\mc{F}}_m}\middle|\mb{W}\right]
=\bE\left[\sum_{m=1}^{M}W_m\left(1-p_{m,K}\right)^{N_m}\middle|\mb{W}\right] \nonumber\\
\stackrel{(a)}{\leq}&\bE\left[\sum_{m=1}^{M}W_m\left(1-p_{K}^{\min}\right)^{N_m}\middle|\mb{W}\right] \nonumber\\
\stackrel{(b)}{\leq}&\bE\left[\sum_{m=1}^{M}W_m\left(1-p_{K}^{\min}\right)^{W_{m}/w_{\max}}\middle|\mb{W}\right]\nonumber\\
\stackrel{(c)}{=}&\bE\Bigg[\sum_{m=1}^{M}W_m\left(1-p_{K}^{\min}\right)^{W_{m}/w_{\max}}\mathbbm{1}_{\{W_m\leq\ol{w}_m\}}\nonumber\\
&\quad+\sum_{m=1}^{M}W_m\left(1-p_{K}^{\min}\right)^{W_{m}/w_{\max}}\mathbbm{1}_{\{W_m>\ol{w}_m\}}\bigg|\mb{W}\Bigg] \nonumber\\
\stackrel{(d)}{\leq}&\bE\left[\sum_{m=1}^{M}W_m\mathbbm{1}_{\{W_m\leq\ol{w}_m\}}+\sum_{m=1}^{M}\mathbbm{1}_{\{W_m>\ol{w}_m\}}\middle|\mb{W}\right]\stackrel{(e)}{\leq} K_2,
\end{align}
where step $(a)$ is true for $p_K^{\min}\triangleq\min_{m}p_{m,K}>0$; $(b)$ is true for $w_{\max}=\lceil F_{\max}/c_{\max}\rceil$ and follows from the fact that given the workload $W_m$, the number of flows at AP $m$ is at least $W_m/w_{max}$ (i.e., $N_m\geq W_m/w_{\max}$); $(c)$ is true for some constant $\ol{w}_m>0$ such that $W_m(1-p_{K}^{\min})^{W_m/w_{\max}}\leq1$ holds whenever $W_m>\ol{w}_m$ (since $\lim_{W_m\rightarrow\infty}W_m(1-p_{K}^{\min})^{W_m/w_{\max}}=0$); $(d)$ uses the fact that $(1-p_{K}^{\min})^{W_m/w_{\max}}\leq1$ and the definition of $\ol{w}_m$; $(e)$ is true for $K_2\triangleq\sum_{m=1}^{M}(\ol{w}_m+1)$.

By combining \eqref{eqn:prop:throughput:temp}, \eqref{eqn:prop:throughput:temp1} and \eqref{eqn:prop:throughput:temp2}, and substituting it into \eqref{eqn:prop:throughput:quadractic}, we have 
\begin{align}
\label{eqn:prop:throughput:quadractic:final}
\bE\left[\Delta L(\mb{W})\middle|\mb{W}\right]\leq&-\frac{2\epsilon}{M}\|\mb{W}\|_{1}+K_1+2K_2\nonumber\\
\leq&-\frac{2\epsilon}{M}\|\mb{W}\|+K_1+2K_2
\end{align}
where the last step uses the fact that $\|\mb{x}\|_1\geq\|\mb{x}\|$ for any vector $\mb{x}$.

By substituting \eqref{eqn:prop:throughput:quadractic:final} into \eqref{eqn:prop:throughput:main}, we have 
\begin{align}
\label{eqn:prop:throughput:main:final}
\bE\left[\Delta V(\mb{W})\middle|\mb{W}\right]\leq-\frac{\epsilon}{M}+\frac{K_1+2K_2}{2V(\mb{W})}.
\end{align}
This implies that when $V(\mb{W})$ is sufficiently large, its conditional expected drift is strictly negative. 

Next, we will show that the drift of $V(\mb{W})$ is also bounded, which together with \eqref{eqn:prop:throughput:main:final} establishes the desired result by \cite[Theorem 2.3]{haj82}.

\begin{align}
\label{eqn:prop:throughput:main:BoundDrift}
|\Delta V(\mb{W})|=&\left|\|\mb{W}[t+1]\|-\|\mb{W}[t]\|\right|\mathbbm{1}_{\{\mb{W}[t]=\mb{W}\}} \nonumber\\
\stackrel{(a)}{\leq}&\|\mb{W}[t+1]-\mb{W}[t]\|\mathbbm{1}_{\{\mb{W}[t]=\mb{W}\}} \nonumber\\
\stackrel{(b)}{\leq}&\|\mb{W}[t+1]-\mb{W}[t]\|_{1}\mathbbm{1}_{\{\mb{W}[t]=\mb{W}\}} \nonumber\\
\leq&M\max_{m}|W_m[t+1]-W_m[t]|\mathbbm{1}_{\{\mb{W}[t]=\mb{W}\}} \nonumber\\
\leq&M(\nu_{\max}+1),
\end{align}
where step $(a)$ follows from the triangle inequality for vectors $\mb{x}$ and $\mb{y}$, i.e., $|\|\mb{x}\|-\|\mb{y}\||\leq\|\mb{x}-\mb{y}\|$; $(b)$ uses the fact that $\|\mb{x}\|\leq\|\mb{x}\|_{1}$ for any vector $\mb{x}$.

\section{Heavy-Traffic Analysis}
\label{sec:heavy-traffic}

In this section, we provide a proof of Proposition \ref{prop:JLW:upperBound}. In particular, we show that the proposed JLW Algorithm minimizes the expected workload in the heavy-traffic regime. The proof includes two parts: 1) showing state-space collapse; 2) using the state-space collapse result to obtain an upper bound on the mean workload. Yet, it is worth noting that each flow faces an independent channel fading and may have different service rate and thus its evolution is quite different from traditional FCFS queueing systems. Thus, it calls for a novel technique to establish heavy-traffic optimality of the JLW Algorithm. 

\subsection{State-Space Collapse}

In this subsection, we establish a state-space collapse result under the JLW Algorithm. That is, we develop the upper bound for the deviation of steady-state workloads from their average. This state-space collapse happens because JLW routes each arrival to the AP with the smaller workload to balance the workload across all APs. 

Let $\{\mb{W}^{(\epsilon)}[t]\}_{t\geq0}$ be the workload process under the JLW Algorithm, where we recall that the heavy-traffic parameter $\epsilon$ characterizes the closeness of the traffic intensity $\rho$ and the boundary of the capacity region $\Lambda$, i.e., $\epsilon=M-\rho^{(\epsilon)}>0$. Proposition \ref{prop:JLW:throughput} shows that all moments of the steady-state workload exist. To that end, we use $\wt{\mb{W}}^{(\epsilon)}$ to denote the steady-state workload random vector. Note that the JLW Algorithm tries to equalize the workload across APs, and thus we expect the state space collapses along the direction of a unit vector, all of whose components are equal, i.e., $\mb{c}\triangleq(1/\sqrt{M})_{m=1}^{M}$. Note that $\mb{W}^{(\epsilon)}[t]\Rightarrow\wt{\mb{W}}^{(\epsilon)}$ due to Proposition \ref{prop:JLW:throughput}, where $\Rightarrow$ denotes convergence in distribution. Then, by the continuous mapping theorem, we have 
$\mb{W}_{\parallel}^{(\epsilon)}[t]\Rightarrow\wt{\mb{W}}_{\parallel}^{(\epsilon)}, \text{ and } \mb{W}_{\perp}^{(\epsilon)}[t]\Rightarrow\wt{\mb{W}}_{\perp}^{(\epsilon)}$,
where the projection and the perpendicular vector of any given $M-$dimensional vector $\mb{I}=(I_m)_{m=1}^{M}$ with respect to the vector $\mb{c}$ are defined as follows:
\begin{align*}
\mb{I}_{\parallel}\triangleq\langle\mb{I},\mb{c}\rangle\mb{c}=\frac{I_{\Sigma}}{M}\mb{1},  \text{ and }\mb{I}_{\perp}\triangleq\mb{I}-\mb{I}_{\parallel}=\left(I_m-\frac{I_{\Sigma}}{M}\right)_{m=1}^{M},
\end{align*}
respectively, and $I_{\Sigma}\triangleq\sum_{m=1}^{M}I_m$, where $\mb{1}$ is $M-$dimensional vector of ones.

Next, we will show that under the JLW Algorithm, all moments of $\wt{\mb{W}}_{\perp}^{(\epsilon)}$ are bounded by some constants independent of heavy-traffic parameter $\epsilon>0$.
\begin{proposition} 
\label{prop:SSC}
For any $\delta\in(0,1/(2w))$, under the JLW Algorithm, there exists a sequence of finite positive numbers $\{H_n\}_{n=1,2,\ldots}$ such that 
\begin{align}
\bE\left[\|\wt{\mb{W}}_{\perp}^{(\epsilon)}\|^{n}\right]\leq H_n, \forall n=1,2,\ldots
\end{align}
for all $\epsilon\in(0,M/2)$, where we recall that $w$ is the mean workload of a newly arriving flow.
\end{proposition}
\begin{IEEEproof}
In the following proof, we will omit $\epsilon$ associated with the workload processes for ease of exposition. We consider the Lyapunov function 
$V_{\perp}(\mb{W})\triangleq\|\mb{W}_{\perp}\|$, and its drift is defined as 
\begin{align}
\Delta V_{\perp}(\mb{W})\triangleq \left(V_{\perp}(\mb{W}[t+1])-V_{\perp}(\mb{W}[t])\right)\mathbbm{1}_{\{\mb{W}[t]=\mb{W}\}}.
\end{align}
Since the workload process $\{\mb{W}[t]\}_{t\geq0}$ has both bounded increments and decrements, we can show that the drift of $V_{\perp}(\mb{W}[t])$ is absolutely bounded by some positive constant for all workload vector $\mb{W}$. Indeed, we have 
\begin{align*}
&\left|\Delta V_{\perp}(\mb{W})\right|=\left|\left\|\mb{W}_{\perp}[t+1]\|-\|\mb{W}_{\perp}[t]\right\|\right|\mathbbm{1}_{\{\mb{W}[t]=\mb{W}\}} \nonumber\\
\stackrel{(a)}{\leq}&\left\|\mb{W}_{\perp}[t+1]-\mb{W}_{\perp}[t]\right\|\mathbbm{1}_{\{\mb{W}[t]=\mb{W}\}}   \nonumber\\
\stackrel{(b)}{=}&\left\|\mb{W}[t+1]-\mb{W}[t]-\left(\mb{W}_{\parallel}[t+1]-\mb{W}_{\parallel}[t]\right)\right\|\mathbbm{1}_{\{\mb{W}[t]=\mb{W}\}} \nonumber\\
\stackrel{(c)}{\leq}&\left(\left\|\mb{W}[t+1]-\mb{W}[t]\right\|+\left\|\left(\mb{W}[t+1]-\mb{W}[t]\right)_{\parallel}\right\|\right)\mathbbm{1}_{\{\mb{W}[t]=\mb{W}\}} \nonumber\\
\stackrel{(d)}{\leq}&2\|\mb{W}[t+1]-\mb{W}[t]\|\mathbbm{1}_{\{\mb{W}[t]=\mb{W}\}}\nonumber\\
\stackrel{(e)}{\leq}&2\|\mb{W}[t+1]-\mb{W}[t]\|_{1}\mathbbm{1}_{\{\mb{W}[t]=\mb{W}\}} \nonumber\\
\leq&2M\max_{m}|W_m[t+1]-W_m[t]|\mathbbm{1}_{\{\mb{W}[t]=\mb{W}\}}\stackrel{(f)}{\leq}2M(\nu_{\max}+1),
\end{align*}
where step $(a)$ uses the triangle inequality for vectors $\mb{x}$ and $\mb{y}$, i.e., $\left|\|\mb{x}\|-\|\mb{y}\|\right|\leq\|\mb{x}-\mb{y}\|$; $(b)$ follows from the definition of $\mb{W}_{\perp}\triangleq\mb{W}-\mb{W}_{\parallel}$; $(c)$ uses the fact that $\|\mb{x}-\mb{y}\|\leq\|\mb{x}\|+\|\mb{y}\|$ for two vectors $\mb{x}$ and $\mb{y}$ and the fact that $\mb{x}_{\parallel}-\mb{y}_{\parallel}=(\mb{x}-\mb{y})_{\parallel}$; $(d)$ uses the fact that $\|\mb{x}_{\parallel}\|\leq\|\mb{x}\|$; $(e)$ uses the fact that $\|\mb{x}\|\leq\|\mb{x}\|_{1}$ for any vector $\mb{x}$; $(f)$ is true for $\nu_{\max}\triangleq A_{\max}\lceil F_{\max}/c_{\max}\rceil$ and follows from \eqref{eqn:dynamic:workload}.

Next, we will show that when $V_{\perp}(\mb{W})$ is sufficiently large, it has a strictly negative drift independent of $\epsilon$. This together with the absolute boundedness of the drift establishes the desired result by \cite[Theorem 2.3]{haj82}. However, it is not easy to directly study the drift of $\|\mb{W}_{\perp}\|$. Instead, it is easier to study the drift of $\|\mb{W}\|^2$ and $\|\mb{W}_{\parallel}\|^2$, which provides a proper upper bound on the drift of $\|\mb{W}_{\perp}\|$. Indeed, 
\begin{align}
\label{eqn:prop:SSC:main}
&\Delta V_{\perp}(\mb{W}) = \left(V_{\perp}(\mb{W}[t+1]) - V_{\perp}(\mb{W}[t])\right)\mathbbm{1}_{\{\mb{W}[t]=\mb{W}\}} \nonumber\\
=&\left(\sqrt{\|\mb{W}_{\perp}[t+1]\|^2} - \sqrt{\|\mb{W}_{\perp}[t]\|^2}\right)\mathbbm{1}_{\{\mb{W}[t]=\mb{W}\}} \nonumber\\
\stackrel{(a)}{\leq}&\frac{1}{2\|\mb{W}_{\perp}[t]\|}\left(\|\mb{W}_{\perp}[t+1]\|^2-\|\mb{W}_{\perp}[t]\|^2\right)\mathbbm{1}_{\{\mb{W}[t]=\mb{W}\}} \nonumber\\
\stackrel{(b)}{=}&\frac{1}{2\|\mb{W}_{\perp}\|}\left(\Delta L(\mb{W})-\Delta L_{\parallel}(\mb{W})\right),
\end{align}
where step $(a)$ follows from the fact that $f(x)=\sqrt{x}$ is concave for $x\geq0$ and thus $f(y)-f(x)\leq f'(x)(y-x)=(y-x)/(2\sqrt{x})$ with $y=\|\mb{W}_{\perp}[t+1]\|^2$ and $x=\|\mb{W}_{\perp}[t]\|^2$; $(b)$ uses the fact that $\|\mb{x}_{\perp}\|^2=\|\mb{x}\|^2-\|\mb{x}_{\parallel}\|^2$ for any vector $\mb{x}$, and is true for $L(\mb{W})\triangleq\|\mb{W}\|^2$, $L_{\parallel}(\mb{W})\triangleq\|\mb{W}_{\parallel}\|^2$, and 
\begin{align}
\Delta L(\mb{W})\triangleq \left(L(\mb{W}[t+1])-L(\mb{W}[t])\right)\mathbbm{1}_{\{\mb{W}[t]=\mb{W}\}}\\
\Delta L_{\parallel}(\mb{W})\triangleq \left(L_{\parallel}(\mb{W}[t+1])-L_{\parallel}(\mb{W}[t])\right)\mathbbm{1}_{\{\mb{W}[t]=\mb{W}\}}.
\end{align}

Next, we consider the conditional expectations of $\Delta L(\mb{W})$ and $\Delta L_{\parallel}(\mb{W})$, respectively. From \eqref{eqn:prop:throughput:quadractic}, \eqref{eqn:prop:throughput:temp}, and \eqref{eqn:prop:throughput:temp2}, we have
\begin{align}
\label{eqn:prop:SSC:Lmain}
\bE\left[\Delta L(\mb{W})\middle|\mb{W}\right]
\leq&2w\langle\mb{W},\bE\left[\mb{A}\middle|\mb{W}\right]-\bs{\lambda}\rangle -\frac{2\epsilon}{M}\|\mb{W}\|_{1} \nonumber\\
& + K_1 + 2K_2 ,
\end{align}
where $K_1$ and $K_2$ are some positive constants.

Next, we consider the term $\langle\mb{W},\bE\left[\mb{A}\middle|\mb{W}\right]-\bs{\lambda}\rangle$ in \eqref{eqn:prop:SSC:Lmain}.
\begin{align}
\label{eqn:prop:SSC:Ltemp1}
\langle\mb{W},\bE\left[\mb{A}\middle|\mb{W}\right]-\bs{\lambda}\rangle\stackrel{(a)}{=}& W_{\min}\bE\left[A_{\Sigma}\middle|\mb{W}\right]-\langle\mb{W},\bs{\lambda}\rangle \nonumber\\
=&W_{\min}\lambda_{\Sigma}-\sum_{m=1}^{M}\lambda_mW_m \nonumber\\
=&-\sum_{m=1}^{M}\lambda_m\left(W_m - W_{\min}\right)\nonumber\\
\stackrel{(b)}{\leq}&-\lambda_{\min}\sum_{m=1}^{M}|W_m-W_{\min}| \nonumber\\
=&-\lambda_{\min}\|\mb{W}-W_{\min}\mb{1}\|_{1} \nonumber\\
\stackrel{(c)}{\leq}&-\lambda_{\min}\|\mb{W}-W_{\min}\mb{1}\| \nonumber\\
\stackrel{(d)}{\leq}&-\lambda_{\min}\left\|\mb{W}-\frac{1}{M}W_{\Sigma}\mb{1}\right\|\nonumber\\
\stackrel{(e)}{\leq}&-\delta\|\mb{W}_{\perp}\|,
\end{align}
where step $(a)$ is true for $W_{\min}\triangleq\min_{m}W_m$ and follows from the definition of the JLW Algorithm; $(b)$ is true for $\lambda_{\min}\triangleq\min_{m}\lambda_m$; $(c)$ follows from the fact that $\|\mb{x}\|_{1}\geq\|\mb{x}\|$ for any vector $\mb{x}$; $(d)$ uses the fact that $W_{\Sigma}/M$ minimizes the convex function $\|\mb{W}-y\mb{1}\|$ over $y\in\mathbb{R}$; $(e)$ is true since $\lambda_{\min}>\delta$ for any $\delta\in(0,1/(2w))$ and $\epsilon\in(0,M/2)$.

By substituting \eqref{eqn:prop:SSC:Ltemp1} into \eqref{eqn:prop:SSC:Lmain}, we have 
\begin{align}
\label{eqn:prop:SSC:Lmain:final}
\bE\left[\Delta L(\mb{W})\middle|\mb{W}\right]\leq-\frac{2\epsilon}{\sqrt{M}}\|\mb{W}_{\parallel}\|-2w\delta\|\mb{W}_{\perp}\|+K_1 + 2K_2.
\end{align}

On the other hand, we have 
\begin{align}
\label{eqn:prop:SSC:Lparallel:final}
&\bE\left[\Delta L_{\parallel}(\mb{W})\middle|\mb{W}\right]=\bE\left[\langle\mb{c},\mb{W}[t+1]\rangle^2-\langle\mb{c},\mb{W}[t]\rangle^2\middle|\mb{W}\right] \nonumber\\
=&\bE\left[\langle\mb{c},\mb{W}+\bs{\nu}-\bs{\mu}\rangle^2-\langle\mb{c},\mb{W}\rangle^2\middle|\mb{W}\right] \nonumber\\
=&\bE\left[2\langle\mb{c},\mb{W}\rangle\langle\mb{c},\bs{\nu}-\bs{\mu}\rangle + \langle\mb{c},\bs{\nu}-\bs{\mu}\rangle^2\middle|\mb{W}\right]\nonumber\\
\geq&2\langle\mb{c},\mb{W}\rangle\langle\mb{c},\bE\left[\bs{\nu}-\bs{\mu}\middle|\mb{W}\right]\rangle\nonumber\\
\stackrel{(a)}{\geq}&2\|\mb{W}_{\parallel}\|\frac{1}{\sqrt{M}}\sum_{m=1}^{M}\left(\bE[\nu_m|\mb{W}]-1\right)\nonumber\\
=&2\|\mb{W}_{\parallel}\|\frac{1}{\sqrt{M}}\left(\bE\left[\nu_{\Sigma}\right]-M\right)\stackrel{(b)}{=}-\frac{2\epsilon}{\sqrt{M}}\|\mb{W}_{\parallel}\|,
\end{align}
where step $(a)$ uses the fact that $\mu_m\leq1, \forall m=1,2,\ldots,M$; $(b)$ follows from the facts that $\bE[\nu_{\Sigma}]=\rho$ and $\epsilon=M-\rho$.

By substituting \eqref{eqn:prop:SSC:Lmain:final} and \eqref{eqn:prop:SSC:Lparallel:final} into \eqref{eqn:prop:SSC:main}, we have 
\begin{align}
\bE\left[\Delta V_{\perp}(\mb{W})\middle|\mb{W}\right]\leq&\frac{1}{2\|\mb{W}_{\perp}\|}\left(-2w\delta\|\mb{W}_{\perp}\|+K_1+2K_2\right)\nonumber\\
=&-w\delta + \frac{K_1+2K_2}{2\|\mb{W}_{\perp}\|}.
\end{align}
Hence, when $V_{\perp}(\mb{W})=\|\mb{W}_{\perp}\|$ is sufficiently large, its expected drift is strictly negative, independent of heavy-traffic parameter $\epsilon$.
\end{IEEEproof}

\subsection{Upper Bound Analysis}
Having established the state-space collapse result, we are ready to provide the upper bound on the mean workload under the JLW Algorithm in the heavy-traffic regime. In Proposition \ref{prop:JLW:throughput}, we have shown that all moments of steady-state workloads are bounded under the JLW Algorithm. This enables us to analyze its heavy-traffic performance by using the methodology of ``setting the drift of a Lyapunov function equal to zero'' (see \cite{erysri12}).

We will omit the superscript $(\epsilon)$ associated with the workload for brevity in the rest of proof. To facilitate the proof, we introduce $U_m\triangleq 1-\mu_m$ and thus the evolution of the workload can be rewritten as  
\begin{align}
\label{eqn:UpperB:evolution}
\mb{W}[t+1] = \mb{W}[t] + \bs{\nu}[t] - \mb{1} + \mb{U}[t],
\end{align}
where $\mb{U}[t]\triangleq(U_{m}[t])_{m=1}^{M}$. Note that $U_m$ is different from the unused service in traditional queues. Indeed, recall that $\mathbbm{1}_{\mc{F}_m}\leq \mu_{m}\leq1$ and thus $0\leq U_m\leq\mathbbm{1}_{\ol{\mc{F}}_m}$, where we recall that $\mc{F}_m$ denotes the event that at least one flow in AP $m$ has the maximum channel rate $c_{\max}$ and $\ol{\mc{F}}_m$ is the complement of the event $\mc{F}_m$. If all flows at AP $m$ do not have the maximum channel rate and the served flow has the size larger than its channel rate, then the workload may not decrease by one (i.e., $\mu_m=0$), which implies that $U_m$ is equal to $1$. However, the flow at AP $m$ does receive the service and does not incur any unused service. This difference causes that the technique in addressing unused service in \cite{erysri12} does not apply and requires additional non-trivial efforts.

In order to derive an upper bound on $\bE\left[\sum_{m=1}^{M}\wt{W}_m\right]$, we need the following fundamental identity \cite[Lemma 8]{erysri12}:
\begin{align}
\label{eqn:UpperB:main}
&\bE\left[\langle\mb{c},\wt{\mb{W}}\rangle\langle\mb{c},\mb{1}-\bs{\nu}\rangle\right]\nonumber\\
=&\frac{1}{2}\bE\left[\langle\mb{c},\bs{\nu}-\mb{1}\rangle^2\right]+\frac{1}{2}\bE\left[\langle\mb{c},\wt{\mb{U}}\rangle^2\right]\nonumber\\
&+\bE\left[\langle\mb{c},\wt{\mb{W}}+\bs{\nu}-\mb{1}\rangle\langle\mb{c},\wt{\mb{U}}\rangle\right],
\end{align}
where $\wt{\mb{U}}$ is a random vector with the same distribution as the steady-state distribution of the process $\{\mb{U}[t]\}_{t\geq0}$. The identity \eqref{eqn:UpperB:main} is derived by setting the expected drift of $\langle\mb{c},\mb{W}\rangle^2$ to $0$ due to the existence of second moment of steady-state workload under the JLW algorithm by Proposition \ref{prop:JLW:throughput}.

First, we consider the left-hand-side (LHS) of \eqref{eqn:UpperB:main}.
\begin{align}
\label{eqn:UpperB:LHS}
\bE\left[\langle\mb{c},\wt{\mb{W}}\rangle\langle\mb{c},\mb{1}-\bs{\nu}\rangle\right]=&\frac{1}{\sqrt{M}}\left(M-\bE\left[\nu_{\Sigma}\right]\right)\bE\left[\langle\mb{c},\wt{\mb{W}}\rangle\right]\nonumber\\
=&\frac{\epsilon}{M}\bE\left[\sum_{m=1}^{M}\wt{W}_m\right].
\end{align}

Next, we will provide an upper bound for each individual term on the right-hand side (RHS) of \eqref{eqn:UpperB:main}. By simply setting the expected drift of $\langle\mb{c},\mb{W}\rangle$ equal to zero, we have 
\begin{align}
\label{eqn:UpperB:unused}
\bE\left[\langle\mb{c},\wt{\mb{U}}\rangle\right]=\bE\left[\langle\mb{c},\mb{1}\rangle-\langle\mb{c},\bs{\nu}\rangle\right]
=&\frac{1}{\sqrt{M}}\left(M-\bE\left[\nu_{\Sigma}\right]\right)\nonumber\\
=&\frac{\epsilon}{\sqrt{M}},
\end{align}
which implies 
\begin{align}
\label{eqn:UpperB:unused1}
\bE\left[\sum_{m=1}^{M}\wt{U}_m\right]=\epsilon.
\end{align}

For the first term on the RHS of \eqref{eqn:UpperB:main}, we have 
\begin{align}
\label{eqn:UpperB:first}
&\frac{1}{2}\bE\left[\langle\mb{c},\bs{\nu}-\mb{1}\rangle^2\right]=\frac{1}{2M}\bE\left[\left(\nu_{\Sigma}-M\right)^2\right] \nonumber\\
=&\frac{1}{2M}\bE\left[\left(\nu_{\Sigma}-\rho-\epsilon\right)^2\right]
=\frac{1}{2M}\left(\text{Var}(\nu_{\Sigma})+\epsilon^2\right).
\end{align}

For the second term on the RHS of \eqref{eqn:UpperB:main}, we have 
\begin{align}
\label{eqn:UpperB:second}
\frac{1}{2}\bE\left[\langle\mb{c},\wt{\mb{U}}\rangle^2\right]\stackrel{(a)}{\leq}&\frac{1}{2}\langle\mb{c},\mb{1}\rangle\bE\left[\langle\mb{c},\wt{\mb{U}}\rangle\right]\stackrel{(b)}{=}\frac{1}{2}\epsilon,
\end{align}
where step $(a)$ follows from the fact that $\wt{U}_m\leq1$, and $(b)$ uses \eqref{eqn:UpperB:unused}.

For the last term on the RHS of \eqref{eqn:UpperB:main}, we have
\begin{align}
\label{eqn:UpperB:third}
&\bE\left[\langle\mb{c},\wt{\mb{W}}+\bs{\nu}-\mb{1}\rangle\langle\mb{c},\wt{\mb{U}}\rangle\right] \nonumber\\
\stackrel{(a)}{=}&\bE\left[\langle\mb{c},\wt{\mb{W}}^{+}\rangle\langle\mb{c},\wt{\mb{U}}\rangle\right]-\bE\left[\langle\mb{c},\wt{\mb{U}}\rangle^2\right]\nonumber\\
\leq&\bE\left[\langle\mb{c},\wt{\mb{W}}^{+}\rangle\langle\mb{c},\wt{\mb{U}}\rangle\right]\nonumber\\
\stackrel{(b)}{=}&\bE\left[\langle\wt{\mb{W}}^{+}_{\parallel},\wt{\mb{U}}_{\parallel}\rangle\right] =\bE\left[\langle\wt{\mb{W}}^{+}-\wt{\mb{W}}^{+}_{\perp},\wt{\mb{U}}-\wt{\mb{U}}_{\perp}\rangle\right]\nonumber\\
=&\bE\left[\langle\wt{\mb{W}}^{+},\wt{\mb{U}}\rangle+\langle\wt{\mb{W}}^{+}_{\perp},\wt{\mb{U}}_{\perp}\rangle-\langle\wt{\mb{W}}^{+},\wt{\mb{U}}_{\perp}\rangle-\langle\wt{\mb{W}}_{\perp}^{+},\wt{\mb{U}}\rangle\right]\nonumber\\
\stackrel{(c)}{=}&\bE\left[\langle\wt{\mb{W}}^{+},\wt{\mb{U}}\rangle\right]+\bE\left[\langle-\wt{\mb{W}}^{+}_{\perp},\wt{\mb{U}}\rangle\right],
\end{align}
where step $(a)$ is true for vector $\mb{I}^{+}$ denoting $\mb{I}[t+1]$ and follows the evolution of the workload $\mb{W}[t]$ (cf. \eqref{eqn:UpperB:evolution}); $(b)$ follows from the fact that $\wt{\mb{W}}_{\parallel}^{+}$ and $\wt{\mb{U}}_{\parallel}$ are along the same direction $\mb{c}$; $(c)$ uses the fact that $\langle\wt{\mb{W}}^{+},\wt{\mb{U}}_{\perp}\rangle=\langle\wt{\mb{W}}^{+}_{\perp},\wt{\mb{U}}_{\perp}\rangle+\langle\wt{\mb{W}}^{+}_{\parallel},\wt{\mb{U}}_{\perp}\rangle=\langle\wt{\mb{W}}^{+}_{\perp},\wt{\mb{U}}_{\perp}\rangle$.

Next, we consider terms in the RHS of \eqref{eqn:UpperB:third}. For the term $\bE\left[\langle\wt{\mb{W}}^{+},\wt{\mb{U}}\rangle\right]$, we have 
\begin{align}
\label{eqn:UpperB:third:temp1}
&\bE\left[\langle\wt{\mb{W}}^{+},\wt{\mb{U}}\rangle\right]=\bE\left[\sum_{m=1}^{M}\wt{W}_m[t+1]\wt{U}_m[t]\right]\nonumber\\
\leq&\bE\left[\sum_{m=1}^{M}\left(\wt{W}_m[t]+\nu_m[t]\right)\wt{U}_m[t]\right]\nonumber\\
\stackrel{(a)}{\leq}&\bE\left[\sum_{m=1}^{M}\wt{W}_m\wt{U}_m\right]+\sqrt{\bE\left[\sum_{m=1}^{M}\nu_m^2\right]\bE\left[\sum_{m=1}^{M}\wt{U}_m^2\right]}\nonumber\\
\stackrel{(b)}{\leq}&\bE\left[\sum_{m=1}^{M}\wt{W}_m\wt{U}_m\right]+\sqrt{\bE\left[\left(\sum_{m=1}^{M}\nu_m\right)^2\right]\bE\left[\sum_{m=1}^{M}\wt{U}_m\right]}\nonumber\\
\stackrel{(c)}{=}&\bE\left[\sum_{m=1}^{M}\wt{W}_m\wt{U}_m\right]+\sqrt{\epsilon\bE\left[\nu_{\Sigma}^2\right]},
\end{align}
where step $(a)$ follows from Cauchy$-$Schwarz inequality; $(b)$ uses the fact that $\wt{U}_m\leq1$; $(c)$ uses \eqref{eqn:UpperB:unused1}.

For the term $\bE\left[\sum_{m=1}^{M}\wt{W}_m\wt{U}_m\right]$ in \eqref{eqn:UpperB:third:temp1}, we have
\begin{align}
\label{eqn:UpperB:third:tempIm}
&\bE\left[\sum_{m=1}^{M}\wt{W}_m\wt{U}_m\right]
\stackrel{(a)}{\leq}\bE\left[\sum_{m=1}^{M}\wt{W}_m\left(1-p_{m,K}\right)^{\wt{N}_{m}}\right]\nonumber\\
=&\bE\left[\sum_{m=1}^{M}\wt{W}_m\left(1-p_{m,K}\right)^{\frac{\wt{N}_{m}}{2}}\left(1-p_{m,K}\right)^{\frac{\wt{N}_{m}}{2}}\right] \nonumber\\
\stackrel{(b)}{\leq}&\bE\left[\sum_{m=1}^{M}\wt{W}_m\left(1-p_K^{\min}\right)^{\frac{\wt{W}_m}{2w_{max}}}\left(1-p_{m,K}\right)^{\frac{\wt{N}_{m}}{2}}\right] \nonumber\\
\stackrel{(c)}{\leq}&\left(\hat{w}_{\max}+1\right)\bE\left[\sum_{m=1}^{M}\left(1-p_{m,K}\right)^{\frac{\wt{N}_{m}}{2}}\right]\nonumber\\
\stackrel{(d)}{\leq}&\left(\hat{w}_{\max}+1\right)\left(\bE\left[\sum_{m=1}^{M}\left(1-p_{m,K}\right)^{d\wt{N}_{m}}\right]\right)^{\frac{1}{2d}}M^{\frac{2d-1}{2d}}\nonumber\\
\stackrel{(e)}{\leq}&\left(\hat{w}_{\max}+1\right)\left(\bE\left[\sum_{m=1}^{M}(p_{m,0})^{\wt{N}_{m}}\right]\right)^{\frac{1}{2d}}M^{\frac{2d-1}{2d}}\nonumber\\
\stackrel{(f)}{\leq}&\left(\hat{w}_{\max}+1\right)\left(\bE\left[\sum_{m=1}^{M}\wt{U}_m\right]\right)^{\frac{1}{2d}}M^{\frac{2d-1}{2d}}\nonumber\\
\stackrel{(g)}{=}&\left(\hat{w}_{\max}+1\right)\epsilon^{\frac{1}{2d}}M^{\frac{2d-1}{2d}},
\end{align}
where step $(a)$ uses the fact that $U_m=1-\mu_m\leq\mathbbm{1}_{\ol{\mc{F}}_m}$; $(b)$ is true for $p_K^{\min}\triangleq\min_{m}p_{m,K}>0$ and $w_{\max}=\lceil F_{\max}/c_{\max}\rceil$ and the fact that the number of flows at AP $m$ is at least $\wt{W}_m/w_{\max}$ (i.e., $\wt{N}_{m}\geq W_m/w_{\max}$); $(c)$ is true for $\hat{w}_{\max}\triangleq\max_{m=1,2,\ldots,M}\hat{w}_{m}$ and $\hat{w}_m$ is some positive constant such that $\wt{W}_m(1-p_K^{\min})^{\wt{W}_m/(2w_{\max})}\leq1$ whenever $\wt{W}_m>\hat{w}_m$ (since $\lim_{\wt{W}_m\rightarrow\infty}\wt{W}_m(1-p_K^{\min})^{\wt{W}_m/(2w_{\max})}=0$), and is derived using similar steps in \eqref{eqn:prop:throughput:temp2}; $(d)$ is true for some $d>1$ such that $(1-p_{m,K})^d\leq p_{m,0}$ (which is the possible due to the assumptions that $p_{m,0}>0$ and $p_{m,K}>0$) and follows from H\"{o}lder's inequality; $(e)$ uses the definition of the constant $d$; $(f)$ uses the fact that $U_m\geq\mathbbm{1}_{\mc{G}_m}$ and $\mc{G}_m$ denotes the event that all flows at AP $m$ do not have available channels, i.e., $\mu_m=0$; $(g)$ uses \eqref{eqn:UpperB:unused1}.
 
By substituting \eqref{eqn:UpperB:third:tempIm} into \eqref{eqn:UpperB:third:temp1}, we have 
\begin{align}
\label{eqn:UpperB:third:1}
\bE\left[\langle\wt{\mb{W}}^{+},\wt{\mb{U}}\rangle\right]\leq\left(\hat{w}_{\max}+1\right)\epsilon^{\frac{1}{2d}}M^{\frac{2d-1}{2d}}+\sqrt{\epsilon\bE\left[\nu_{\Sigma}^2\right]}.
\end{align}

For the term $\bE\left[\langle-\wt{\mb{W}}^{+}_{\perp},\wt{\mb{U}}\rangle\right]$ in \eqref{eqn:UpperB:third}, we have
\begin{align}
\label{eqn:UpperB:third:2}
\bE\left[\langle-\wt{\mb{W}}^{+}_{\perp},\wt{\mb{U}}\rangle\right]\stackrel{(a)}{\leq}&\sqrt{\bE\left[\|\wt{\mb{W}}_{\perp}^{+}\|^2\right]\bE\left[\|\wt{\mb{U}}\|^2\right]}\nonumber\\
\stackrel{(b)}{\leq}&\sqrt{\bE\left[\|\wt{\mb{W}}_{\perp}\|^2\right]\bE\left[\sum_{m=1}^{M}\wt{U}_m\right]} \nonumber\\
\stackrel{(c)}{\leq}&\sqrt{H_2\epsilon},
\end{align}
where step $(a)$ uses Cauchy$-$Schwarz inequality; $(b)$ uses the fact that $U_m\leq1$; $(c)$ uses the state-space collapse result (cf. Proposition \ref{prop:SSC}) and \eqref{eqn:UpperB:unused1}.

By substituting \eqref{eqn:UpperB:third:1} and \eqref{eqn:UpperB:third:2} into \eqref{eqn:UpperB:third}, we have 
\begin{align}
\label{eqn:UpperB:third:final}
&\bE\left[\langle\mb{c},\wt{\mb{W}}+\bs{\nu}-\mb{1}\rangle\langle\mb{c},\wt{\mb{U}}\rangle\right] \nonumber\\
\leq&\left(\hat{w}_{\max}+1\right)\epsilon^{\frac{1}{2d}}M^{\frac{2d-1}{2d}}+\sqrt{\epsilon\bE\left[\nu_{\Sigma}^2\right]}+\sqrt{H_2\epsilon}\triangleq G(\epsilon).
\end{align}

By substituting \eqref{eqn:UpperB:LHS}, \eqref{eqn:UpperB:first}, \eqref{eqn:UpperB:second}, and \eqref{eqn:UpperB:third:final} into \eqref{eqn:UpperB:main}, we have 
\begin{align*}
\epsilon\bE\left[\sum_{m=1}^{M}\wt{W}_m\right]\leq\frac{1}{2}\left(\text{Var}(\nu_{\Sigma})+\epsilon^2\right)+\frac{1}{2}M\epsilon+MG(\epsilon),
\end{align*}
which implies the desired result as $\epsilon\downarrow0$.

\section{Conclusions}

In this paper, we studied the optimal load-balancing design in high-density wireless networks with both channel fading and flow-level dynamics. We discussed the performance deficiencies of existing policies and developed a workload-aware load-balancing scheme in the presence of dynamic flows. We showed that our proposed load-balancing algorithm not only achieves maximum system throughput, but also minimizes the mean total workload in heavy-traffic regimes. In addition, our analysis implies that the mean total workload performance under our proposed algorithm is robust to the number of APs, which is strongly desirable in high-density wireless networks. Finally, extensive simulations were performed to confirm our theoretical results.

\appendices


\section{Characterization of Capacity Region}
\label{APP:capacity}
%
(1) (Necessity) Assume that $\rho>M$ is true. Consider the Lyapunov function $J(\mb{W})\triangleq\sum_{m=1}^{M}W_m$. Then, we have 
\begin{align}
&\bE\left[J(\mb{W}[t+1])-J(\mb{W}[t])\middle|\mb{W}[t]=\mb{W}\right] \nonumber\\
\stackrel{(a)}{=}&\sum_{m=1}^{M}\bE\left[\nu_m[t]-\mu_m[t]\middle|\mb{W}[t]=\mb{W}\right] \nonumber\\
=&\bE\left[\nu_{\Sigma}[t]\right]-\sum_{m=1}^{M}\bE\left[\mu_m[t]\middle|\mb{W}[t]=\mb{W}\right] \nonumber\\
\stackrel{(b)}{\geq}&\rho-M\stackrel{(c)}{>}0,
\end{align}
where step $(a)$ uses the dynamic of workload (cf. \eqref{eqn:dynamic:workload}); $(b)$ uses $\nu_{\Sigma}[t]=\sum_{m=1}^{M}\nu_m[t]$, $\bE\left[\nu_{\Sigma}[t]\right]=\rho$, and the fact that $\mu_m[t]\leq1$; $(c)$ uses our contradictory assumption. 

Thus, by \cite[Theorem 3.3.10]{srikant2013communication}, no policy can stabilize the system.

(2) (Sufficiency) Proposition \ref{prop:JLW:throughput} in Section \ref{sec:JLW} shows that any arrival traffic intensity $\rho$ strictly inside $\Lambda$ (i.e., $\rho<M$) can be supported by the policy proposed in Section \ref{sec:JLW}. This together with the necessity proof establishes the desired result.
%

\begin{spacing}{}
\bibliographystyle{abbrv}
\bibliographystyle{IEEEtran}
\bibliography{refs}
\end{spacing}

\end{document}